\documentclass[fleqn,usenatbib]{mnras}

\usepackage{newtxtext,newtxmath}

\usepackage[T1]{fontenc}

\DeclareRobustCommand{\VAN}[3]{#2}
\let\VANthebibliography\thebibliography
\def\thebibliography{\DeclareRobustCommand{\VAN}[3]{##3}\VANthebibliography}

\usepackage{graphicx}	
\usepackage{amsmath}	
\usepackage{csvsimple}
\usepackage{caption}
\usepackage{booktabs}
\usepackage[flushleft]{threeparttable}
\usepackage{hyperref}
\usepackage{comment}
\usepackage[inline]{enumitem}
\usepackage[normalem]{ulem}
\usepackage[table]{xcolor}

\graphicspath{{./}{figures/}}

\title[Gaseous Satellites]{Census of Gaseous Satellites around Local Spiral Galaxies}

\author[Zhu $\&$ Putman]{
Jingyao Zhu,$^{1}$\thanks{E-mail: jingyao.zhu@columbia.edu}
Mary E. Putman,$^{1}$
\\
$^{1}$Department
of Astronomy, Columbia University, New York, NY 10027, USA\\
}

\date{Accepted 2023 March 1. Received 2023 February 28; in original form 2022 August 2}

\pubyear{2023}

\begin{document}
\label{firstpage}
\pagerange{\pageref{firstpage}--\pageref{lastpage}}
\maketitle

\begin{abstract}
We present a search for gas-containing dwarf galaxies as satellite systems around nearby spiral galaxies using 21 cm neutral hydrogen (HI) data from the Arecibo Legacy Fast ALFA (ALFALFA) Survey. We have identified 15 spiral `primary' galaxies in a local volume of 10 Mpc with a range of total masses, and have found 19 gas-containing dwarf satellite candidates within the primaries' virial volumes ($R_{200}$) and 46 candidates within $2R_{200}$. Our sensitivity using ALFALFA data converts to $M_{\rm HI} \approx 7.4 \times 10^{6}$ $M_{\odot}$ at 10 Mpc, which includes 13 of the 26 gaseous dwarf galaxies in the Local Group, and the HI properties of our sample are overall similar to these 13.  We found $0-3$ gaseous satellites per host galaxy within $R_{200}$ and $0-5$ within $2R_{200}$, which agrees with the low numbers present for the Milky Way and M31.  There is also agreement with the star-forming satellite numbers per host in the deep optical surveys SAGA and ELVES, and the Auriga cosmological simulations. When scaled to $R_{200}$, the optical surveys do not show a trend of increasing quenched fraction with host mass; there is a slight increase in the total number of gaseous satellites with host mass for our sample.  The low numbers of gaseous/star-forming satellites around spiral hosts are consistent with the idea that a universal and effective satellite quenching mechanism, such as ram pressure stripping by the host halo, is likely at play.
\end{abstract}




\section{Introduction} \label{sec:intro}
Galaxy stellar mass ($M_{*}$) and environment are found to be two major factors in determining a galaxy's star formation stage \citep{kauffmann_environmental_2004,baldry_galaxy_2006,peng_mass_2010}. For central galaxies (the most massive galaxies within a halo), the cold gas removal and quenching of star formation are sensitive to the stellar mass and internal feedback processes \citep{croton_many_2006,dalla_vecchia_simulating_2008}.  For satellite galaxies of a more massive primary, environmental quenching -- the decrease in star formation rate due to interactions with the primary's halo, becomes important \citep{peng_mass_2012,wetzel_galaxy_2012,brown_cold_2017}. Environmental quenching is more effective for the low mass satellites \citep{bluck_how_2020-1}, especially the dwarf galaxy satellites \citep{boselli_origin_2008,boselli_cold_2014,hester_ic_2010}. 
Dwarf galaxies that are satellites of a larger galaxy are more likely to be quenched, unlike their field counterparts, which are predominately gas-rich \citep{geha_stellar_2012,phillips_dichotomy_2014,karachentsev_morphological_2018}. In the observed Local Group dwarf galaxies \citep{mcconnachie_observed_2012,simon_faintest_2019-1}\footnote{For an updating version of the \citet{mcconnachie_observed_2012} data, see \url{http://www.astro.uvic.ca/~alan/Nearby_Dwarf_Database.html}}, this bias of satellite galaxies being quenched is particularly strong (e.g., \citealt{grcevich_h_2009,2015ApJ...808L..27W}). A correlation has been identified between the lack of gas content and the proximity to the primaries \citep{grcevich_h_2009,spekkens_dearth_2014}, and potentially with the proximity to the Local Group halo \citep{putman_gas_2021}. The dependence for satellite quenching on the distance to a primary suggests that the Milky Way (MW) and M31's halo environments play an important role in removing the satellites' star formation fuel.

Several possible mechanisms contribute to a dwarf satellite galaxy's cold gas removal and are not easily separable (for a recent observational review, see \citealt{cortese_dawes_2021}). Ram pressure stripping (RPS), the direct removal of a satellite’s ISM via an interaction with a hot halo \citep{gunn_infall_1972-1}, is an effective quenching mechanism across various mass scales and environments \citep{roediger_ram_2005,tonnesen_star_2012,bekki_galactic_2014,steyrleithner_effect_2020}. RPS accounts for over $50 \%$ of the quenched satellites in the Auriga cosmological simulations of Milky Way analogs \citep{grand_auriga_2017-1,simpson_quenching_2018}.
RPS is more effective at closer distances to the host, where the host halo density and the satellite velocity are both higher \citep{gunn_infall_1972-1}, and thus can naturally account for the observed distance dependence in satellite quenching.
Abundant observational evidence confirms that spiral galaxy halos are filled with diffuse halo gas that will play a role in stripping satellite galaxies of gas as they move through it (e.g., \citealt{bowen_structure_2016,qu_hstcos_2019,tchernyshyov_cgm2_2022}).
RPS is likely accompanied by gravitational effects like tidal stirring \citep{mayer_tidal_2001,mayer_simultaneous_2006,boselli_origin_2014} and tidal interactions with the host \citep{mateo_velocity_2008,simpson_quenching_2018}, though the latter requires very close orbital encounters, where RPS typically already removed most of the gas \citep{bahe_star_2015,cora_semi-analytic_2018,cortese_dawes_2021}. Other than the environmental quenching mechanisms above, cosmic reionization (e.g., \citealt{babul_dwarf_1992,gnedin_cosmological_2000,somerville_can_2002}) and stellar and supernova feedback (e.g., \citealt{mac_low_control_2004,mckee_theory_2007,agertz_toward_2013}) both contribute to the gas removal but are likely insufficient for the complete quenching of typical dwarf galaxies \citep{weisz_star_2014,emerick_gas_2016,rodriguez_wimberly_suppression_2019}.

Is the Local Group unique in its dwarf galaxy population and gas-quenching mechanisms?  In recent years, a growing number of optical studies have characterized the satellite systems around nearby ($<10-20$ Mpc) primaries beyond the Local Group.  These studies primarily focus on MW-like primaries \citep{spencer_survey_2014,smercina_lonely_2018,kondapally_faint_2018,tanaka_missing_2018,smercina_saga_2020}, but have also included a more diverse sample of host galaxies \citep{cohen_dragonfly_2018,carlsten_wide-field_2020,carlsten_radial_2020,carlsten_luminosity_2021,carlsten_exploration_2022-1}. At larger distances, $D \in (20,40)$ Mpc, the ongoing SAGA survey \citep{geha_saga_2017,mao_saga_2021} is building towards a one-hundred MW-analog sample and mapping the satellite systems around them. 
A missing link in these studies is the gaseous satellite population for galaxies beyond the Local Group to determine if the mechanisms of cold gas removal are universal. In particular, the SAGA survey finds a much lower ($\sim 15 \%$) quenched fraction, which is in tension with the mostly gas-quenched Local Group picture.

This study presents a search for dwarf satellite systems around nearby spiral galaxies in the local volume ($D < 10$ Mpc) using the Arecibo Legacy Fast ALFA (ALFALFA) HI data \citep{haynes_arecibo_2011,haynes_arecibo_2018}. The structure of this paper is summarized as follows. In \S\ref{sec:data}, we introduce the 21 cm HI data we use in this study and plan to use in the future. Then, we describe the selection methods of the primary spiral galaxies in \S \ref{sec:methods_spiral_host_definition} and the satellite candidate galaxies in \S \ref{subsec:sat_search}. We assess the Milky Way and M31's detectable gaseous satellites by mock external observations of the Local Group dwarf galaxies in \S \ref{subsec:lg_dwarf_population}. The results are shown in \S \ref{sec:results}, focusing on the gaseous satellite properties in \S \ref{subsec:satellite_properties} and the number per spiral host in \S \ref{subsec:satellite_number}. This is followed by a direct comparison with recent deep optical surveys: ELVES and SAGA (\S \ref{sec:elves_saga_discussion}), in terms of the star-forming/gaseous satellite abundances (\S\ref{subsec:sf_sats_elves_saga}) and the quenched fraction (\S \ref{subsec:quenched_fraction}). Finally, we highlight our key findings and provide an outlook for future work in \S\ref{sec:summary}.

\section{Data} \label{sec:data}

We use the complete ALFALFA Extragalactic Source Catalog ($\alpha.100$, \citealt{haynes_arecibo_2011,haynes_arecibo_2018}) as our data source for the primary and satellite galaxy search. The ALFALFA survey has $\sim 7000$ deg$^{2}$ of sky coverage, a $4 \arcmin$ circular beam size at the full width at half maximum (FWHM), and a spectral resolution smoothed to $10$ km $\rm s^{-1}$ \citep{giovanelli_arecibo_2005,haynes_arecibo_2018}. Its typical rms noise level of $\sim 2.4$ mJy per beam is deeper than previous wide-field HI surveys.
The ALFALFA complete catalog $\alpha.100$ has $\sim 31,500$ sources with 21 cm HI emission and has been recently cross-matched with the Sloan Digital Sky Survey (SDSS) results for optical counterparts (ALFALFA-SDSS; \citealt{durbala_alfalfa-sdss_2020-1}).

The ALFALFA survey's depth is a primary advantage over previous HI surveys\footnote{For comparison of major `blind' HI surveys, see \url{http://egg.astro.cornell.edu/alfalfa/science.php}.}, as the detectability of low-HI mass sources is crucial to our study. The HI detectability (in terms of HI mass) can be expressed as,

\begin{equation}\label{eq:completeness_theory}
    M_{\rm HI}(M_{\odot}) = 2.356 \times 10^{5} \times D_{\rm Mpc}^{2} \times \frac{W_{50}}{\sqrt{W_{50}/(2 \cdot V_{\rm res})}} \times \frac{5 \cdot \sigma_{\rm 21cm}}{1000},
\end{equation}

where $W_{50}$ is the FWHM of the 21 cm emission signal, $V_{\rm res}$ is the spectral resolution, and $\sigma_{\rm 21cm}$ (in mJy) is the survey's rms noise level per beam. Equation \ref{eq:completeness_theory} shows the $5 \sigma$ HI mass detectability limit given a constant rms beam noise $\sigma_{\rm 21cm}$ \citep{giovanelli_arecibo_2005}. With ALFALFA, we can release the constant beam noise assumption and adopt the actual data sensitivity \citep[see][equations 4 and 5]{haynes_arecibo_2011}. 

We use the state-of-the-art ALFALFA data in this paper; in a future paper, we will follow up this work using the ongoing Australian Square Kilometer Array Pathfinder HI All-Sky Survey (WALLABY; \citealt{koribalski_wallaby_2020}). WALLABY has a spectral resolution of $V_{\rm res}=4$ km $\rm s^{-1}$, a targeted noise of $\sigma_{\rm 21cm} = 1.6$ mJy per 30$\arcsec$ beam, and a $\sim 14439$ deg$^{2}$ coverage across the southern sky. Upon completion, WALLABY will capture a larger (deeper) HI sample than ALFALFA (equation \ref{eq:completeness_theory}). The current WALLABY pilot data target three $\sim 60$ deg$^{2}$ fields of galaxy groups/clusters at a median redshift distance $D \approx 60$ Mpc \citep{westmeier_wallaby_2022}; these fields have very few spiral galaxies within 10 Mpc. Below, we predict the WALLABY capabilities in detecting gaseous dwarf satellites compared to ALFALFA, and we leave the detailed WALLABY analyses to future papers when new data becomes available.

\begin{figure}
    \centering
    \includegraphics[width=1.0\linewidth]{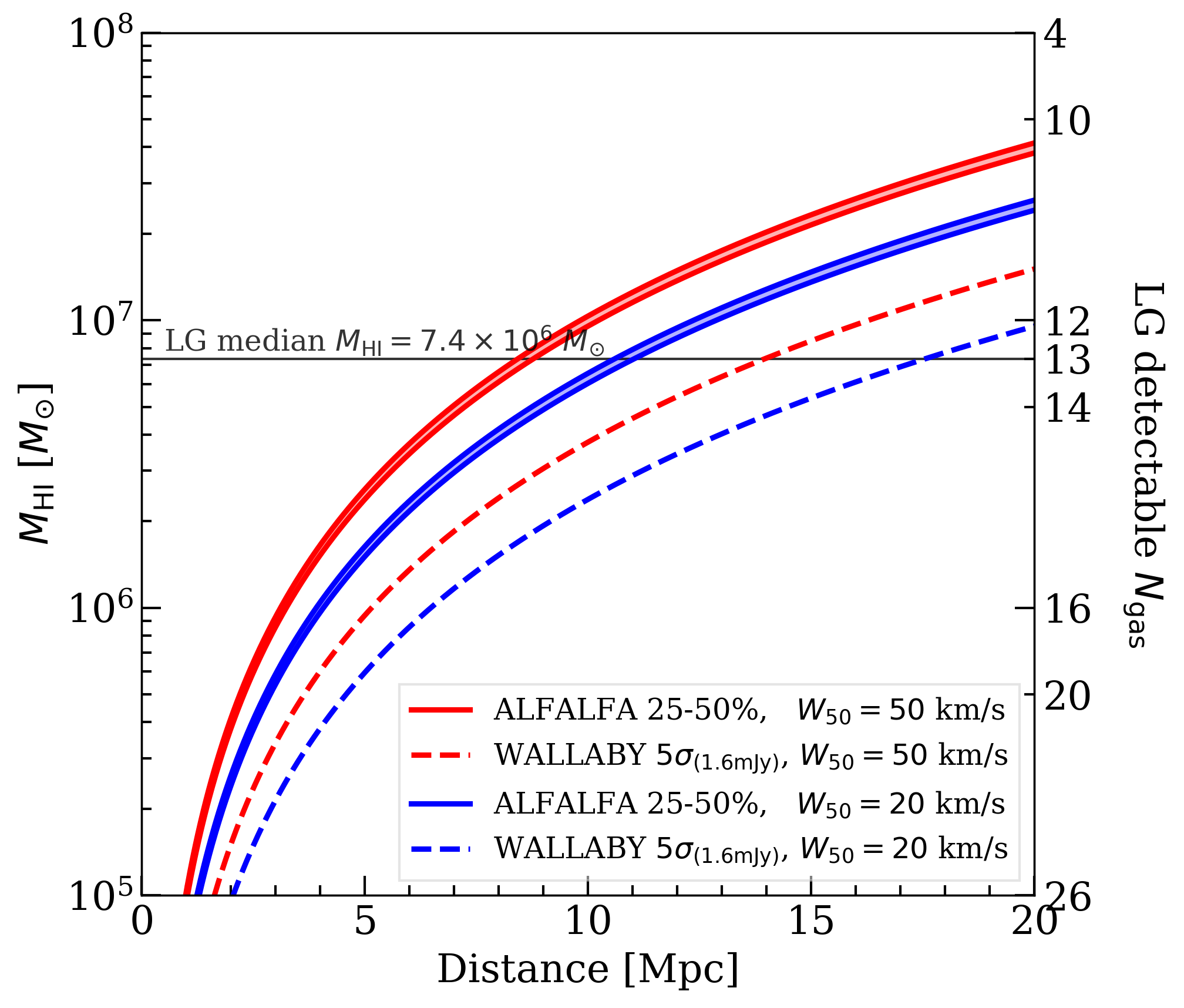}
    \caption{Detectable HI mass versus distance for the ALFALFA and WALLABY surveys. Solid curves with shadows are ALFALFA `Code 1' ($S/N > 6.5$) sensitivities at $25 - 50 \%$ confidence limits \citep{haynes_arecibo_2011}. Dashed curves are the $5 \sigma$ HI mass thresholds adopting the targeted WALLABY sensitivity $\sigma_{\rm 21cm}=1.6$ mJy beam$^{-1}$ (equation \ref{eq:completeness_theory}). The right-hand y-axis gives the detectable number of the Local Group (LG) gas-containing dwarfs (out of 26 total) at the detectable HI mass on the left-hand y-axis. The horizontal line annotates the median HI mass of the LG gaseous dwarfs. The two sampled velocity widths, $W_{50} = 50, 20$ km $\rm s^{-1}$, are typical values to the gaseous LG dwarfs ($W_{50, \rm median} \approx 32$ km $\rm s^{-1}$, see Figure \ref{fig:mhi_census} below).}
    \label{fig:survey_sensitivity_with_count}
\end{figure}

Figure \ref{fig:survey_sensitivity_with_count} summarizes the HI depth versus source distance for the ALFALFA survey (this paper) and the ongoing WALLABY survey at full sensitivity (future follow-up work). To put HI depths in the context of Local Group dwarf galaxies, we refer to the recent sample of \citet{putman_gas_2021} (119 dwarf galaxies within $2$ Mpc, $26$ of which HI gas-containing). The y-axis on the right of Figure \ref{fig:survey_sensitivity_with_count} shows the number of gas-containing Local Group (LG) dwarfs detectable at the HI mass limit shown on the left.
We chose to place the detectability cut at the LG dwarf median HI mass of $7.4 \times 10^{6}$ $M_{\odot}$ (Figure \ref{fig:survey_sensitivity_with_count}), which converts to a $10$-Mpc local volume of the Milky Way, where $\gtrsim 50\%$ of the LG-like gaseous dwarfs are detectable by ALFALFA. The 10-Mpc choice is a balance between having a larger host sample and being able to detect lower HI-mass satellites. In our future WALLABY work, the distance cut can be extended to a few Mpc deeper for this HI mass cut (dashed curves in Figure \ref{fig:survey_sensitivity_with_count}).

\section{Methods}\label{sec:methods}
\subsection{Local Spiral Galaxy Selection}\label{sec:methods_spiral_host_definition}
We select the primary spiral galaxies (`host') sample under the initial motivation of finding Milky Way analogs in the local universe. Within the ALFALFA catalog, we intend to search for spiral galaxies with masses comparable to the MW and without severe tidal disturbances by close companions. As discussed in \S\ref{sec:data}, we also constrain our spiral host search between the ALFALFA catalog distances: $2 < D_{\alpha} < 10$ Mpc, to enable the detectability of at least $\sim 50 \%$ of the low-mass, gas-containing dwarf galaxy satellites beyond the Local Group. To select masses comparable to the MW, we apply a broad filter on the HI FWHM ($W_{50}$) between $150$ and $600$ km $\rm s^{-1}$ to include Milky Way-like galaxies with a broad range of inclinations. The distance $D_{\alpha} \in (2,10)$ Mpc and velocity width $W_{50} \in [150, 600]$ km $\rm s^{-1}$ initial filters resulted in 18 ALFALFA local spiral galaxies. We then examined the optical-band morphology of these galaxies to ensure they are clear of major tidal disturbances. This excluded NGC 672 as part of an interacting dwarf pair \citep{pearson_local_2016}; and NGC 4747, which is in significant interaction with a close companion, NGC 4725 \citep{haynes_neutral_1979}. Furthermore, we excluded NGC 5014 from our sample as 
its projected virial radius ($R_{200}$, as outlined later in \S \ref{sec:methods_spiral_host_definition}) has only partial ALFALFA coverage, and it belongs to a loose group with NGC 5005 and NGC 5033 \citep{noordermeer_westerbork_2005}, where both group members are spiral galaxies brighter than NGC 5014 and outside of the ALFALFA field.

\begin{figure}
    \centering
    \includegraphics[width=1.0\linewidth]{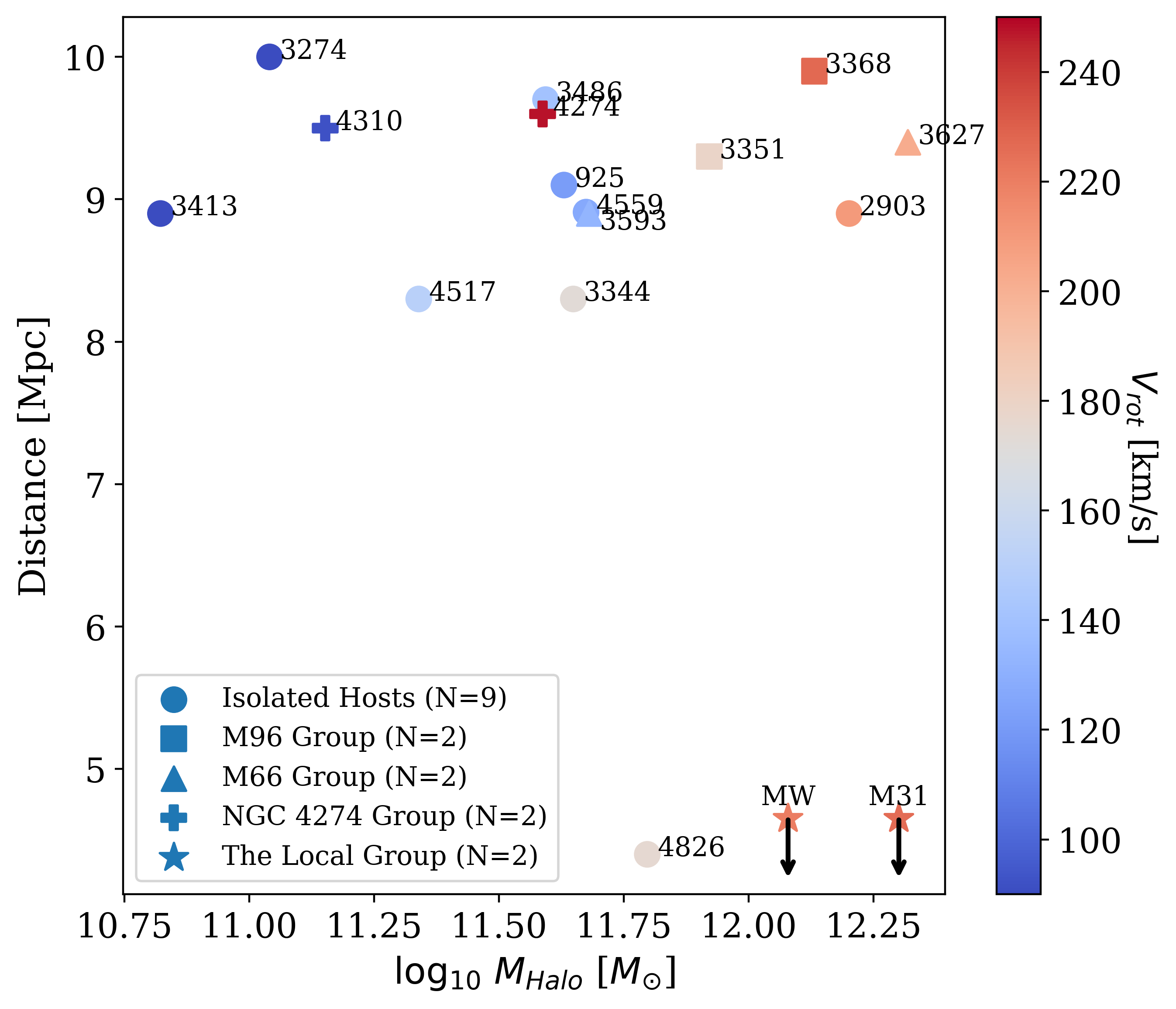}
    \caption{Distance versus halo mass for the ALFALFA spiral host sample, colored by the derived rotation velocities and annotated with the NGC numbers. We obtain the halo masses from abundance matching, adopting stellar mass values from literature (Table \ref{table:analog_properties}). For the MW and M31, we consistently adopt $M_{\rm Halo} \approx 1.2 \times 10^{12}$ $M_{\odot}$ and $2.0 \times 10^{12}$ $M_{\odot}$; $V_{\rm rot} \approx 220$ km $\rm s^{-1}$ and $226$ km $\rm s^{-1}$, respectively. We annotate the three host pairs in our sample that have overlapping projected virial volumes (Figure \ref{fig:virgo_field_sky_distribution}) and are frequently associated as galaxy group members; see Appendix \ref{sec:results_individual_hosts}. The high rotation velocity with the intermediate halo mass of NGC 4274 is likely due to an underestimation of stellar mass (and thus halo mass) from the distance uncertainty, see \S \ref{subsec:NGC4274_ComaI}.}
    \label{fig:analog_mass_stats}
\end{figure}

The resulting host sample has 15 local spiral galaxies, and we summarize their physical properties in Table \ref{table:analog_properties} and Figure \ref{fig:analog_mass_stats}. To estimate the dark matter halo masses, we adopt literature stellar masses $M_{*}$ and apply stellar-to-halo mass (SHM) abundance matching \citep{moster_constraints_2010}. In addition to abundance matching, we also estimate the hosts' inclination-corrected rotation velocities, $V_{\rm rot} = W_{50}/(2 \cdot \sin(i))$, where $i$ is the inclination angle from The Catalog and Atlas of the LV galaxies (LVG; \citealt{karachentsev_updated_2013}) where available. For the three galaxies without LVG inclinations, we use \citet{truong_high-resolution_2017} for NGC 4310 and 
\citet{paturel_hyperleda_2003} for NGC 3486 and 4274.

To compare our host sample with the Milky Way and Andromeda, we first acknowledge that the two Local Group spirals have a range of literature masses; see, e.g., \citet{bland-hawthorn_galaxy_2016-1} for a review. In this work, we consistently adopt $1.2 \times 10^{12}$ $M_{\odot}$ for MW and $2.0 \times 10^{12}$ $M_{\odot}$ for M31. It follows that our host sample is relatively low-mass in comparison, with 12 of the 15 primaries being less massive than the Milky Way. In addition, \citet{mcgaugh_dark_2021} finds that abundance matching techniques tend to slightly overestimate halo masses, which places our host sample further towards the low-mass end. The low masses result from our broad $W_{50}$ selection for host galaxies, which is motivated to select Milky Way-like galaxies even if they are close to face-on. The three primaries with $V_{\rm rot} \approx 90$ km $\rm s^{-1}$ are clearly not MW analogs, where one (NGC 4310) forms a small group with a larger spiral (NGC 4274), and the other two (NGC 3413, 3274) are relatively isolated. We keep them in the primary sample to explore the satellites around smaller spiral galaxies question.

The large scatter in host masses (Table \ref{table:analog_properties}, Figure \ref{fig:analog_mass_stats}) motivates an examination of the physical halo sizes, characterized by the virial radii ($R_{200}$). 
Given a halo mass $M_{\rm Halo}$ from abundance matching, the virial radius is given by $M_{\rm Halo} = \frac{4\pi}{3} (200 \cdot \rho_{\rm crit,0}) R_{200}^{3}$, where $\rho_{\rm crit,0}$ is the present day critical density (Planck Collaboration \citeyear{collaboration_planck_2020}). When searching for satellites, we refer to the $R_{200}$ values in Table \ref{table:analog_properties} to characterize the primaries' halo regions  (\S\ref{subsec:sat_search}).
\begin{table*}
    \begin{threeparttable}
    \caption{Properties of the local spiral host galaxy sample}
    \label{table:analog_properties}
\begin{tabular}{ccccccccccc}
\hline
Name & RA & DEC & Distance$^{(2)}$ & $V_{\rm rot}^{(3)}$ & $\log M_{\rm HI}^{(4)}$ & $\log M_{*}$ & $M_{*}$ ref$^{(5)}$ & $\log M_{\rm Halo}$ & $R_{200}$ & $V_{\rm esc}^{(6)}$ \\
NGC$^{(1)}$ & (deg) & (deg) & (Mpc) & (km $\cdot \rm s^{-1}$) & ($M_{\odot}$) & ($M_{\odot}$) &   & ($M_{\odot}$) & (kpc) & (km $\cdot \rm s^{-1}$)\\
\hline
3413	&	162.844	&	32.763	&	8.9	&	90	&	8.45	&	8.44	&	SDSS	&	10.8	&	86	&   82\\
3274	&	158.075	&	27.668	&	10$^{(7)}$	&	90	&	9.07	&	8.88	&	Spitzer	&	11.0	&	101 &   97\\
4310	&	185.609	&	29.204	&	9.5	&	92	&	7.46	&	9.10	&	SDSS	&	11.2	&	110	&   105\\
4517	&	188.186	&	0.113	&	8.3	&	150	&	9.32	&	9.46	&	SDSS	&	11.3	&	127$^{(8)}$	&   122\\
4274	&	184.961	&	29.617	&	9.6	&	248	&	8.2	&	9.90	&	SDSS	&	11.6	&	154	&   147\\
3486	&	165.1	&	28.973	&	9.7	&	140	&	9.41	&	9.91	&	Spitzer	&	11.6	&	155	&   148\\
925	&	36.808$^{(9)}$	&	33.578	&	9.1	&	122	&	9.71	&	9.97	&	IR	&	11.6	&	159	&   152\\
3344	&	160.879	&	24.921	&	8.3$^{(7)}$	&	173	&	9.45	&	10.00	&	Spitzer	&	11.7	&	161	&   154\\
4559	&	188.995	&	27.958	&	8.9$^{(7)}$	&	127	&	9.69	&	10.04	&	IR	&	11.7	&	164	&   157\\
3593	&	168.65	&	12.814	&	8.9	&	133	&	8.33	&	10.05	&	Spitzer	&	11.7	&	165	&   158\\
4826 (M64)	&	194.204	&	21.66	&	4.4	&	175	&	8.26	&	10.22	&	Spitzer	&	11.8	&	181	&   173\\
3351 (M95)	&	160.99	&	11.704	&	9.3	&	179	&	9.03	&	10.38	&	Spitzer	&	11.9	&	199 &   190\\
3368 (M96)	&	161.689	&	11.822	&	9.9	&	227	&	9.26	&	10.60	&	Spitzer	&	12.1	&	234	&   223\\
2903	&	143.044	&	21.508	&	8.9	&	210	&	9.56	&	10.66	&	Spitzer	&	12.2	&	246	&   236\\
3627 (M66)	&	170.063	&	12.988	&	9.4	&	202	&	8.92	&	10.75	&	Spitzer	&	12.3	&	270 &   258 \\
\hline
\end{tabular}
\begin{tablenotes}
      \small
      \item (1). Sorted by stellar mass in ascending order. Messier identification is provided where available. (2). Distances are ALFALFA catalog distances $D_{\alpha}$ except for three galaxies where more updated resolved stellar population distances are adopted, see note 7 and \S\ref{sec:results_individual_hosts} for details. (3). Rotation velocities are derived from HI velocity widths $W_{50}/(2\sin (i))$, adopting inclination angles $i$ from LVG \citep{karachentsev_updated_2013} wherever available. For NGC 3486 and 4274 we reference the HyperLEDA values \citep{paturel_hyperleda_2003}, and for NGC 4310 we adopt an updated value from \citet{truong_high-resolution_2017}. (4). The HI masses are ALFALFA catalog values and updated for the cases with different distances, see note 7. (5). The stellar masses are literature values scaled to our adopted distances in column 2. The reference keys are: 'IR' \citep{de_los_reyes_revisiting_2019}, 'Spitzer' \citep{cook_spitzer_2014}, and 'SDSS' (the ALFALFA-SDSS catalog; \citealt{durbala_alfalfa-sdss_2020-1}). (6). Point-mass ($M_{\rm Halo}$) escape velocities evaluated at hosts' $R_{200}$. (7). The ALFALFA distances $D_{\alpha}$ for NGC 3274 and 3344 are replaced by \citet{sabbi_resolved_2018} TRGB distances, and for NGC 4559 is replaced by \citet{2017AJ....154...51M} TRGB distance, to which the relevant masses are scaled. (8). The ALFALFA survey covers two sky areas, Virgo ($\sim 7.5$h to $\sim 16.5$h R.A.) and Pisces ($\sim 22$h to $\sim 3$h R.A.), both with $0{\degr} < \delta < 36{\degr}$ declination range. NGC 4517 is locating at the declination boundary, whose projected $R_{200}$ halo region is partially covered in ALFALFA. (9). NGC 925 is the only spiral host in our sample within the Pisces field (see note 8), see the separated R.A. range in Figure \ref{fig:virgo_field_sky_distribution}.
    \end{tablenotes}

    \end{threeparttable}
\end{table*}

\subsection{Gaseous Satellite Search}\label{subsec:sat_search}
We determine the satellite-host association by combining an on-sky spatial cut and a spectroscopic velocity cut, as frequently adopted in previous studies \citep{mcconnachie_observed_2012,geha_saga_2017,mao_saga_2021}. This combination enables effective characterization of a host's halo region without requiring two observational challenges: accurate line-of-sight distances and proper motions. Unlike the on-sky/projected component, the line-of-sight distance component is often missing for dwarf galaxies \citep{shaya_action_2017,kourkchi_cosmicflows-4_2020} and can reach Mpc-scale uncertainties even with targeted deep optical imaging \citep{cohen_dragonfly_2018,carlsten_luminosity_2021,carlsten_exploration_2022-1}. In ALFALFA, the catalog distances ($D_{\alpha}$) are literature values when available and otherwise group- or velocity-flow- model values, associated with distance uncertainties $\sigma_{D} = 2.0 \pm 1.2$ Mpc for $D_{\alpha} < 10$ Mpc \citep{jones_alfalfa_2018,haynes_arecibo_2018}. These uncertainties exceed the size of a Milky Way-like or smaller halo (see $R_{200}$ in Table \ref{table:analog_properties}), so $D_{\alpha}$ is insufficient to determine satellite association along the line-of-sight direction. We use the $D_{\alpha}$ distance as a supplement to our associations based on spectroscopic velocities and projected distances, as described in this section.

We employ two complementary spatial cuts (projected distance $d_{\rm proj}$) and two spectroscopic velocity\footnote{Uncertainties in the ALFALFA spectroscopic velocities are relatively negligible because of the high spectral resolution (\S \ref{sec:data}).} cuts (velocity difference between satellite and host, $\delta V$) as satellite definitions.  The more restrictive option aims for a higher confidence of satellite association, while the other aims for a larger, more complete sample. The two projected distance filters are:

\begin{enumerate}
    \item $d_{\rm proj} \leq R_{200}$. Selecting satellites within the hosts' projected virial volumes, see Table \ref{table:analog_properties} for $R_{200}$.
    \item $d_{\rm proj} \leq 2R_{200}$. Extending out to $2R_{200}$, capturing a larger projected halo area, with possibly a higher degree of field contamination.
\end{enumerate}
Combined with two velocity filters,
\begin{enumerate}
    \item $|\delta V| \coloneqq |(V_{\rm sat}-V_{\rm host})|  \leq V_{\rm esc}(R_{200})$, thereafter `$V_{\rm esc}$'. The escape velocity $V_{\rm esc}(R_{200})$ (see Table \ref{table:analog_properties}) is evaluated at the host's virial radius assuming a point-mass model ($\sqrt{\frac{2GM_{\rm Halo}}{R_{200}}}$). It is an indicator of gravitationally bound conditions.
    \item $|\delta V| \leq 300$ km $\rm s^{-1}$. A uniform $300$ km $\rm s^{-1}$ cut is consistently more inclusive than $V_{\rm esc}$; it is an upper limit motivated criterion commonly used in the literature. We denote the satellite candidates with $V_{\rm esc} < |\delta V| \leq 300$ km $\rm s^{-1}$ as `$V_{300}$' below.
\end{enumerate}

The ALFALFA primary sample spans a range of masses (and hence halo sizes, gravitational potential depths), where many are less massive than the Milky Way (Table \ref{table:analog_properties}). 
The large variations suggest that a physical cut (e.g., $R_{200}$ and $V_{\rm esc}$) is better than a uniform cut. For the projected distance filters, instead of using the commonly adopted literature value 300 kpc (Milky Way-like virial radius), we supply the two options, $R_{200}$ for higher sample confidence, and $2R_{200}$ for larger coverage and completeness. When later comparing with other Milky Way-analog studies that adopted 300 kpc, we scale them to $R_{200}$ and $2R_{200}$ accordingly (see Figures \ref{fig:count_comprehensive}, \ref{fig:Mg_cut_satellite_count_Rvir_comprehensive}, \S \ref{subsec:sf_sats_elves_saga}). Similarly, we use $V_{\rm esc}$ as the major velocity selection, and $300$ km $\rm s^{-1}$ (an empirical velocity upper limit for Milky Way-like satellites, see \citealt{mcconnachie_observed_2012,geha_saga_2017,mao_saga_2021}) as an upper limit. The $300$ km $\rm s^{-1}$ option aims at including higher-velocity satellites while accounting for the uncertainties in $V_{\rm esc}$ and in the point-mass gravity model (e.g., for hosts in groups/pairs). Satellite candidates with $V_{\rm esc} < |\delta V| \leq 300$ km $\rm s^{-1}$ ($V_{300}$) are likely more contaminated and are denoted as upper error bars in our key results later (Figures \ref{fig:count_comprehensive}, \ref{fig:alfa_elves_host_direct_comparison}, and \ref{fig:Mg_cut_satellite_count_Rvir_comprehensive}).

Our ALFALFA dwarf satellite search pipeline is as follows. To identify the low-mass objects, we first apply an HI velocity width filter $W_{50} < 150$ km $\rm s^{-1}$ as a broad selection to encompass all galaxies below the velocity width of our primaries. Around each host's projected field, we generate samples of satellite candidates using each spatial cut combined with each velocity cut defined above. 
We then exclude the known HI gas clouds with no stellar components in the Leo Ring and Leo I Group, i.e., in the field of M66 and M96  (\citealt{stierwalt_arecibo_2009}; thereafter S09). Because the broad $W_{50}$ range potentially includes high $V_{\rm rot}$ objects with small inclinations, we individually examine the massive satellite candidates' optical images and exclude those with $V_{\rm rot} > 90$ km s$^{-1}$. This avoids overlap with our primary sample (Table \ref{table:analog_properties}) and also puts the satellite sample in consistency with the LG dwarf sample, where up to LMC-like satellites are included, and M33 is excluded.

To supplement our satellite selection criteria, we calculate the likelihood the candidates fall within the hosts' three-dimensional (3D) proximity using the ALFALFA catalog distances and uncertainties \citep{jones_alfalfa_2018,haynes_arecibo_2018}. This is done using,
\begin{equation}\label{eqn:Fd3d}
    F_{\rm d_{\rm 3D}} \coloneqq P(\rm d_{3D} \leq 1 Mpc),
\end{equation}
where $F_{\rm d_{\rm 3D}}$ is the likelihood of 3D proximity, i.e., the cumulative probability that the 3D distance between the satellite and the host ($d_{\rm 3D}$) is closer than a certain threshold, here chosen to be $1$ Mpc. This factor compares the relative confidence in 3D proximity among the satellite candidates (tabulated in Table \ref{table:2R200_satellites}, also see Figure \ref{fig:dvel_dproj_scatter} below). We describe the algorithm in more detail in Appendix \ref{appendix:Fd3d} and discuss its implication for individual satellite candidates of interest in Appendix \ref{sec:results_individual_hosts}.


When a satellite candidate falls within more than one host galaxy's projected halo, 
we distinguish which host the satellite is more likely associated with based on the observational proximity. We define a dimensionless separation $\Theta$ between the satellite-host pair: 
\begin{equation} \label{eqn:dimensionless-separation}
    \Theta(\rm sat,host) \coloneqq \sqrt{(d_{\rm proj}/R_{200})^{2} + (\delta V/V_{\rm esc})^{2}},
\end{equation}
where $d_{\rm proj}/R_{200}$ quantifies a dimensionless spatial separation, and $\delta V/V_{\rm esc}$ quantifies a kinematic separation, both normalizing over the host halo size. All three host pairs in our sample with overlapping $R_{200}$ regions (see Figures \ref{fig:analog_mass_stats} and \ref{fig:virgo_field_sky_distribution}), namely the Leo I (NGC 3351 and 3368), M66 (NGC 3593 and 3627), and NGC 4274 (NGC 4274 and 4310) groups, share multiple satellite candidates in their projected $2R_{200}$. We assign the shared satellite candidates to their closer host, characterized by a lower dimensionless separation $\Theta$. It is also used in the mock Local Group observations (\S \ref{subsec:lg_dwarf_population}) and the comparison with the gaseous/star-forming satellites in deep optical surveys later (Figure \ref{fig:alfa_elves_host_direct_comparison}).

\subsection{External Observations of the Gaseous Satellites Around the Milky Way and Andromeda} \label{subsec:lg_dwarf_population}

Extragalactic observations of dwarf galaxies as satellites inevitably face problems of incompleteness and projection effects. The incompleteness results from missing the low-mass end of gaseous dwarf galaxies due to the distance-dependent HI mass sensitivity (Figure \ref{fig:survey_sensitivity_with_count}). The projection effects result from potentially including field galaxies that fall into the projected sky. The Local Group dwarf galaxy sample is relatively complete and has a high distance accuracy. To directly compare the satellite samples of the Milky Way and M31 with our ALFALFA satellite sample, we completed mock external observations of the Local Group at typical ALFALFA analog distances using the satellite selection criteria outlined in \S \ref{subsec:sat_search}.

Taking the full sample of gas-containing dwarf galaxies within 2 Mpc from \citet{putman_gas_2021}, we conducted Monte Carlo mock ALFALFA observations of the Local Group from randomized viewing angles. We experimented with three cases, $D_{\rm mock} = 4.4$, $8.9$, and $10$ Mpc, representing the ALFALFA analog minimum, average, and maximum distances (Table \ref{table:analog_properties}). For each case, we place the observer at $D_{\rm mock}$ away from the Local Group's center of mass, adopting $M_{\rm Halo,MW} = 1.2 \times 10^{12}$ $M_{\odot}$, $R_{200, \rm MW} \approx 224$ kpc and $M_{\rm Halo,M31} = 2.0 \times 10^{12}$ $M_{\odot}$, $R_{200, M31} \approx 266$ kpc, in consistency with \citet{putman_gas_2021}. We then sample $N=10^{5}$ random viewing angles, determine the ALFALFA $25\%$ detectability (\S \ref{sec:data} and \citealt{haynes_arecibo_2011}), and record the resulting gaseous satellites within the Milky Way's and M31's projected $R_{200}$, $2R_{200}$. Similar to \S \ref{subsec:sat_search}, if a satellite falls into both the Milky Way's and M31's projected halos, we assign it to the closer host based on the dimensionless $\Theta$ in equation \ref{eqn:dimensionless-separation}.

Table \ref{table:LG_2Rvir} summarizes the Local Group gas-containing satellites that consistently fall within the projected $2R_{200}$ of the MW and M31, i.e., median $d_{\rm proj} \leq 2R_{200, \rm host}$, in the Monte Carlo mock ALFALFA observations. The satellites' median projected distances range from $70-90\%$ of their 3D distances to the hosts (Table \ref{table:LG_2Rvir}). The HI masses and velocity widths listed in the Table are used in calculating the HI detectability as described in \S \ref{sec:data}. The two distance columns, $D_{\alpha 25}$ and $D_{1.6 \rm mJy}$, show the maximum detectable distances of the dwarf satellites given the HI surveys' mass sensitivities. The resulting Monte Carlo detection frequencies, $\frac{N_{\rm detect}}{N_{\rm tot,MC}}$, are impacted by both the mass sensitivity and the projection angles.

\begin{table*}
\caption{Local Group Gas-containing Dwarf Galaxies within the Spiral Hosts' Projected $2R_{200}$}
\label{table:LG_2Rvir}
\begin{threeparttable}
\begin{tabular}{ccccccccc}
\hline
Name & Host & $d_{\rm 3D,host}^{(1)}$ &   $d_{\rm proj}^{(2)}$  &  $\log M_{\rm HI}$ & $W_{50}^{(3)}$ & $D_{\alpha 25}^{(4)}$ & $D_{1.6 \rm mJy}^{(5)}$ & $(\frac{N_{\rm detect}}{N_{\rm tot,MC}})^{(6)}$\\
 &  & (kpc) &   (kpc) & $(M_{\odot})$ & (km $\rm s^{-1}$) & (Mpc) & (Mpc) & ($\%$)\\
\hline
LMC         & MW  &	51   & 40 $\pm$ 11    & 8.66 &	80$^{\rm (a)}$  &   61.7   &   98.2    & 100  \\
SMC         & MW  &	64   & 45 $\pm$ 16    & 8.66 &	90              &   60.0   &   95.4    & 100  \\
Leo T       & MW  & 409  & 344 $\pm$ 86   & 5.45 &	20$^{\rm (c)}$  &   2.2    &   3.5     &   0  \\ 
Phoenix     & MW  & 415  & 297 $\pm$ 119  & 5.08 &  16$^{\rm (b)}$  &   1.5    &   2.4     &   0  \\

NGC 6822    & MW  & 459  &  412 $\pm$ 91  & 8.11 &    80            &   32.8   &   52.2    &  87  \\ 
\hline
NGC 205     & M31 &	46   &  43 $\pm$ 9     &	5.60 &	38$^{\rm (d)}$ &   2.2    &   3.5  &   0  \\
NGC 185     & M31 &	183  & 164 $\pm$ 37    &	5.04 &	36$^{\rm (d)}$ &   1.2    &   1.9  &   0  \\
IC 10       & M31 &	252  & 179 $\pm$ 74    &	7.70 &	62             &   21.7   &   34.5 & 100  \\
LGS 3       & M31 &	268  & 194 $\pm$ 73    &	5.58 &	21$^{\rm (b)}$ &   2.5    &   3.9  &   0  \\
Pegasus dIrr& M31 & 474   & 370 $\pm$ 106   &	6.77 &	23             &   9.6    &   15.2 &  88  \\
IC 1613     & M31 &	518  & 382 $\pm$ 133   &	7.81 &	27             &   30.5   &   48.4 & 100  \\
\hline
\end{tabular}
\begin{tablenotes}
      \small
      \item (1). The 3D distances to host ($d_{\rm 3D,host}$) are derived from the distances and coordinates of the updated Local Group sample \citep[see][Table 1]{putman_gas_2021}. (2). The projected distance ($d_{\rm proj}$) column lists the median $\pm~1 \sigma$ values from the Monte Carlo mock ALFALFA observations of the Local Group, see \S \ref{subsec:lg_dwarf_population}. (3). $W_{50}$ values are adopting \citet{karachentsev_updated_2013} by default and updated with additional references where more representative gas velocity dispersions are available from, e.g., deep HI observations of the individual dwarf galaxy. References: (a). \citet{staveley-smith_new_2003}; (b). \citet{1997ApJ...490..710Y}; (c). \citet{adams_deep_2018}; (d). \citet{young_neutral_1997}. (4). $D_{\alpha 25}$: maximum detectable distances given the ALFALFA $25\%$ confidence level (Figure \ref{fig:survey_sensitivity_with_count}; \citealt{haynes_arecibo_2011}). (5). $D_{1.6 \rm mJy}$: maximum detectable distances at $5 \sigma$ confidence level assuming WALLABY's targeted rms noise of $\sigma_{\rm 21cm} = 1.6$ mJy (see equation \ref{eq:completeness_theory}). (6). The fractional number of detections ($N_{\rm detect}$) out of the total Monte Carlo tests ($N_{\rm tot,MC}=10^{5}$), i.e., the detection frequencies, where $100 \%$ indicates consistently detected, and $0\%$ never detected. Values shown here are from the mean ALFALFA distance case, $D_{\rm mock}=8.9$ Mpc.
    \end{tablenotes}
    \end{threeparttable}
\end{table*}

From the Monte Carlo tests at the mean ALFALFA host distance ($D_{\rm mock} = 8.9$ Mpc), the median $\pm~34.1$ percentiles total satellite number for the Milky Way's $R_{200}$ is $N=2_{-0}^{+1}$ ($76.6\%$ of the tests return $N=2$, $18.9\%$ return $N=3$, the remaining $N \geq 4$), consisting of the LMC, SMC, and potentially NGC 6822 for some viewing angles. For the Milky Way's $2R_{200}$, this number increases to $N=3_{-0}^{+1}$, with further away galaxies such as Leo A coming into the projected area. Leo T and Phoenix\footnote{Phoenix has confusion with Galactic emission, but the HI velocity and position agree with the stellar component, and hence is included \citep{putman_gas_2021}.} also reside within the Milky Way's projected $2R_{200}$ but are always undetectable ($\frac{N_{\rm detect}}{N_{\rm tot,MC}}=0$) at ALFALFA analog distances, because they are below the mass sensitivity at the mock distances (Table \ref{table:LG_2Rvir}). The Monte Carlo tests at $D_{\rm mock}=8.9$ Mpc give $N=2_{-1}^{+0}$ within M31's $R_{200}$ ($47.8\%$ of the tests return $N=1$, $40.7\%$ return $N=2$, the remaining $N \geq 3$), consisting of IC 10 and either IC 1613 or Pegasus, depending on the viewing angle. Within M31's $2R_{200}$ the number of satellites is $N=4_{-1}^{+0}$, including WLM with the $R_{200}$ satellites. If we adopt the minimum/maximum ALFALFA distance (4.4/10.0 Mpc) instead of the mean, these detectable numbers will increase/decrease by no more than 1. M31's NGC 185, NGC 205, and LGS 3 are always undetectable at ALFALFA analog distances (Table \ref{table:LG_2Rvir}). We note that all of these undetectable Local Group satellites are somewhat in their own category, with small gas masses and no detectable current star formation.

From the fractional perspective, $3/5$ of the MW and $3/6$ of the M31 gas-containing dwarf satellites within projected $2R_{200}$ are detectable in external, ALFALFA-like observations of the Local Group (Table \ref{table:LG_2Rvir}, where $\frac{N_{\rm detect}}{N_{\rm tot,MC}} \geq 85\%$). This confirms that considering the mass limits and projection effects, our choice of ALFALFA host galaxy distance at $\leq 10$ Mpc captures about $50 \%$ of the Local Group primaries' gaseous satellites (\S\ref{sec:data}). If the Milky Way and M31 have comparable satellite HI properties to the ALFALFA host galaxies, this predicts our ALFALFA gaseous satellite sample to be approximately $50 \%$ complete. When presenting the $2R_{200}$ satellites of the Milky Way and M31 in the figures hereafter, we refer to the full sample in Table \ref{table:LG_2Rvir} when comparing with the entire Local Group population (Figures \ref{fig:mhi_w50_scatter_with_sensitivity}, \ref{fig:mhi_census}, \ref{fig:dwarf_galaxy_HI_depletion_lg_comparison}), and to the detectable sample in Table \ref{table:LG_2Rvir} when comparing with extragalactic observations (Figures \ref{fig:dvel_dproj_scatter}, \ref{fig:alfa_elves_saga_luminosity_comparison}). Finally, we note that extragalactic observations may be more contaminated than the mock Local Group observations presented here because we targeted a 2-Mpc volume-confined sample.

\section{Results}\label{sec:results}

\begin{figure*}
    \centering
    \includegraphics[width=0.9\linewidth]{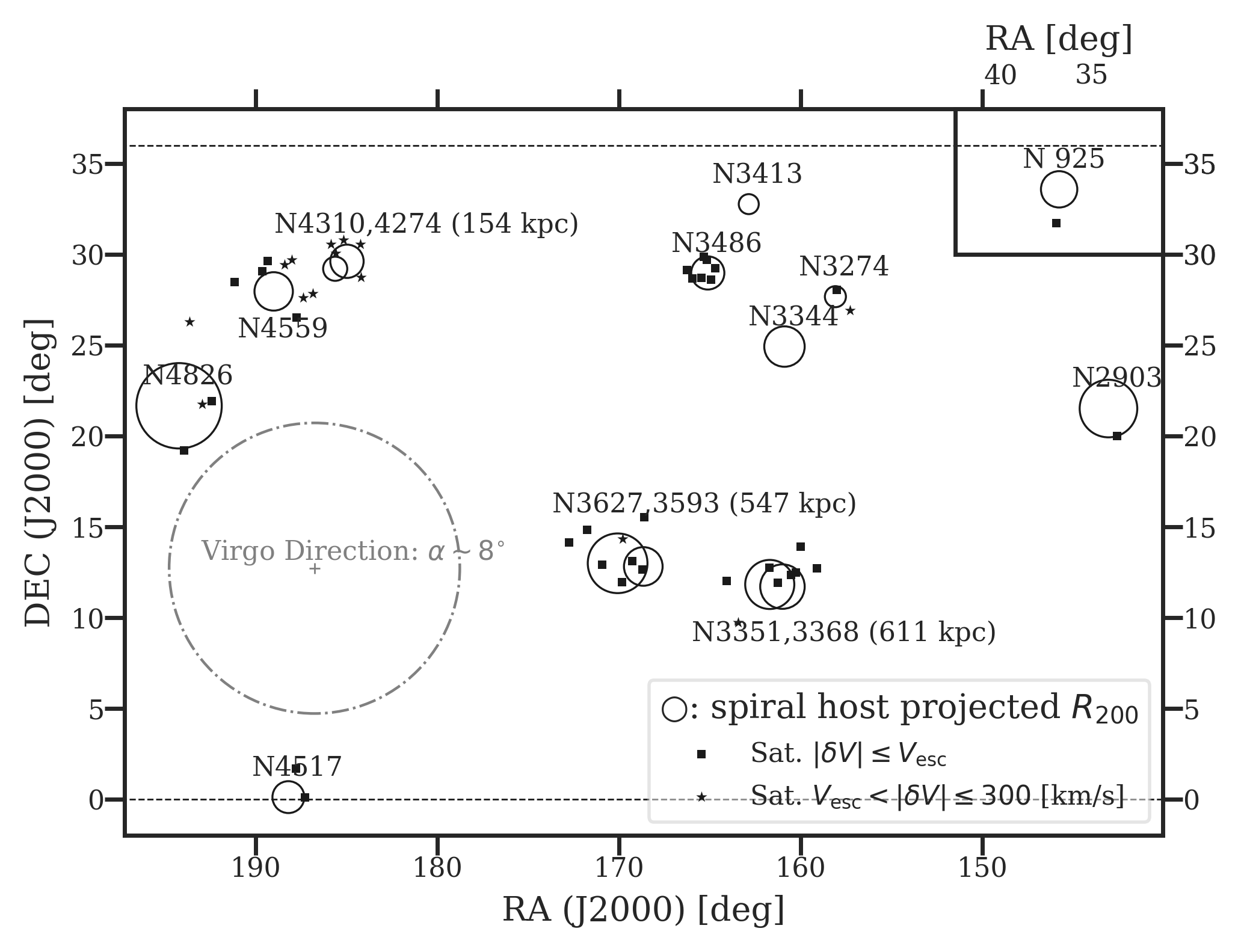} 
    \caption{Sky distribution of the ALFALFA satellite candidates within host projected $2R_{200}$'s. Solid circles are the spiral hosts' projected $R_{200}$ areas. NGC 925 is in the ALFALFA Pisces field (separate RA range; Table \ref{table:analog_properties}), denoted in the upper right corner with a separate RA range, while the DEC range is shared. Squares and stars are the gaseous satellite candidates following the $V_{\rm esc}$, $V_{300}$ velocity criteria (see \S\ref{subsec:sat_search}) within host projected $2R_{200}$ fields. The Virgo cluster direction is denoted with an assigned angular radius of $8{\degr}$. The three host pairs and respective distances from each other ($d_{\rm 3D} \leq 800$ kpc) are annotated; a significant percentage of halo overlap is found for all pairs. NGC 4517 is close to the ALFALFA field edge indicated by the dashed lines (Table \ref{table:analog_properties}) and has incomplete ALFALFA sky coverage.}
    \label{fig:virgo_field_sky_distribution}
\end{figure*}

We present the sky distribution of the ALFALFA satellite candidates within host $2R_{200}$'s in Figure \ref{fig:virgo_field_sky_distribution}. Given the criteria described in \S\ref{subsec:sat_search}, there are 19 satellite candidates within the ALFALFA spiral hosts' virial radii (see the satellite points within the projected $R_{200}$ circles in Figure \ref{fig:virgo_field_sky_distribution}) and 46 candidates in total within the projected $2R_{200}$ areas. Contamination due to the Virgo cluster is likely negligible as it does not overlap any projected $R_{200}$ in our sample (Figure \ref{fig:virgo_field_sky_distribution}, adopting an angular size $\sim 8{\degr}$ for Virgo). The properties of the satellites are described in \S\ref{subsec:satellite_properties}, and the satellite numbers per spiral host are presented in \S\ref{subsec:satellite_number}. The properties of the dwarf satellite candidates are listed in Table \ref{table:2R200_satellites}. 

\begin{table*}
\caption{Properties of the $2R_{200}$ Satellite Candidates Around ALFALFA Spiral Hosts \label{table:2R200_satellites}} 
\begin{threeparttable}
\begin{tabular}{cccccccccccc}
\hline
Host$^{(1)}$ & Sat.$^{(2)}$ & RA$^{(3)}$ & DEC$^{(3)}$ & $D_{\alpha}^{(4)}$ & $V_{\odot}$ & $d_{\rm proj}^{(5)}$ & $\delta V^{(6)}$ & $W_{50}$ & $M_{\rm HI}^{(7)}$ & $F_{\rm d_{\rm 3D}}^{(8)}$ \\
NGC & AGC & (deg) & (deg) & (Mpc) & (km $\rm s^{-1}$) & (kpc) & (km $\rm s^{-1}$) & (km $\rm s^{-1}$) & ($M_{\odot}$) & ($\%$) \\
\hline																					
925	&	1924$^{*}$	&	36.957	&	31.731	&	8.5	&	595	&	294	&	43	&	111	&	1.90e8	&	31.2	\\
\hline																					
2903	&	\textbf{191706}$^{*}$	&	142.555	&	19.992	&	8.2	&	561	&	246	&	4	&	23	&	1.90e7	&	30.1	\\
\hline																					
3274	&	\textbf{731457}$^{*}$	&	157.995	&	28.03	&	11.1	&	454	&	64	&	-87	&	36	&	1.50e7	&	43.5	\\
	&	749315	&	157.284	&	26.904	&	9.2	&	645	&	181	&	104	&	31	&	8.00e6	&	29.2	\\
\hline																					
3351	&	5761$^{*}$	&	159.097	&	12.712	&	11.1	&	604	&	342	&	-173	&	112	&	8.10e7	&	24.1	\\
	&	\textbf{200532}$^{*}$	&	160.503	&	12.333	&	11.1	&	772	&	128	&	-5	&	36	&	2.10e7	&	25	\\
	&	201970$^{*}$	&	161.725	&	12.739	&	11.1	&	636	&	205	&	-141	&	38	&	1.30e7	&	25.5	\\
\hline																					
3368	&	200512$^{*}$	&	159.983	&	13.904	&	11.1	&	1007	&	461	&	114	&	21	&	7.10e6	&	26.3	\\
	&	5812$^{*}$	&	160.237	&	12.473	&	11.1	&	1008	&	270	&	115	&	56	&	4.00e7	&	28.5	\\
	&	\textbf{202024}$^{*}$	&	161.251	&	11.913	&	11.1	&	871	&	76	&	-22	&	24	&	5.30e6	&	30.4	\\
	&	202035$^{*}$	&	164.062	&	12.011	&	11.1	&	989	&	403	&	96	&	30	&	4.30e7	&	28	\\
	&	6014	&	163.425	&	9.729	&	11.1	&	1133	&	467	&	240	&	94	&	7.40e7	&	25.4	\\
\hline																					
3486	&	\textbf{208569}$^{*}$	&	164.695	&	29.22	&	9.8	&	697	&	73	&	20	&	81	&	1.30e7	&	33.5	\\
	&	\textbf{722731}$^{*}$	&	164.933	&	28.607	&	9.8	&	698	&	67	&	21	&	12	&	3.70e6	&	33.5	\\
	&	\textbf{212945}$^{*}$	&	165.165	&	29.708	&	9.4	&	673	&	125	&	-4	&	44	&	3.30e7	&	33.1	\\
	&	\textbf{215232}$^{*}$	&	165.31	&	29.849	&	9.8	&	693	&	151	&	16	&	23	&	1.70e7	&	33.2	\\
	&	\textbf{6102}$^{*}$	&	165.449	&	28.689	&	9.6	&	697	&	71	&	20	&	70	&	2.00e8	&	18.8	\\
	&	\textbf{219369}$^{*}$	&	165.963	&	28.686	&	9.2	&	667	&	137	&	-10	&	22	&	2.00e7	&	33.9	\\
	&	210027$^{*}$	&	166.245	&	29.139	&	11	&	647	&	172	&	-30	&	44	&	2.50e7	&	28.5	\\
\hline																					
3593	&	\textbf{202256}$^{*}$	&	168.689	&	12.65	&	10	&	630	&	26	&	-1	&	42	&	1.30e7	&	33.3	\\
	&	\textbf{210220}$^{*}$	&	169.253	&	13.098	&	10	&	588	&	101	&	-43	&	25	&	1.10e7	&	33.2	\\
\hline																					
3627	&	215282$^{*}$	&	168.613	&	15.534	&	10.9	&	867	&	478	&	151	&	27	&	6.00e6	&	15.9	\\
	&	\textbf{202257}$^{*}$	&	169.81	&	11.953	&	10.4	&	861	&	175	&	145	&	51	&	6.80e7	&	17.6	\\
	&	\textbf{213440}$^{*}$	&	170.908	&	12.896	&	10	&	666	&	136	&	-50	&	22	&	5.20e6	&	34.9	\\
	&	215296$^{*}$	&	171.729	&	14.828	&	11.5	&	913	&	402	&	197	&	44	&	1.20e7	&	20.9	\\
	&	212837$^{*}$	&	172.724	&	14.145	&	10.7	&	880	&	465	&	164	&	22	&	4.10e7	&	26.4	\\
	&	\textbf{215286}	&	169.797	&	14.328	&	10	&	998	&	224	&	282	&	28	&	1.20e7	&	37.4	\\
\hline																					
4274	&	7300	&	184.185	&	28.731	&	9.4	&	1209	&	187	&	292	&	74	&	2.70e8	&	18.9	\\
	&	747848	&	184.234	&	30.553	&	9.8	&	1145	&	189	&	228	&	34	&	2.20e7	&	18.9	\\
	&	220408	&	185.144	&	30.798	&	9.9	&	1101	&	200	&	184	&	38	&	3.20e7	&	18	\\
	&	724774	&	185.844	&	30.566	&	9.8	&	679	&	204	&	-238	&	53	&	2.80e7	&	18.8	\\
	&	\textbf{7438}	&	185.582	&	30.071	&	9.7	&	694	&	118	&	-223	&	101	&	1.60e7	&	19.1	\\
\hline																					
4517	&	225760$^{*}$	&	187.274	&	0.101	&	15.6	&	1198	&	132	&	71	&	25	&	1.50e7	&	0.3$^{\rm (8a)}$	\\
	&	221570$^{*}$	&	187.766	&	1.684	&	16.3	&	1104	&	236	&	-23	&	39	&	4.10e7	&	0.2$^{\rm (8a)}$	\\
\hline																					
4559	&	724906$^{*}$	&	187.736	&	26.512	&	11.6	&	927	&	284	&	114	&	22	&	1.47e7	&	16.7	\\
	&	220847$^{*}$	&	189.309	&	29.627	&	10.3	&	795	&	263	&	-18	&	44	&	1.71e7	&	28.1	\\
	&	724993$^{*}$	&	189.63	&	29.052	&	9.6	&	756	&	191	&	-57	&	37	&	1.88e7	&	30.5	\\
	&	220984$^{*}$	&	191.159	&	28.473	&	12.7	&	945	&	307	&	132	&	50	&	1.21e8	&	8.9	\\
	&	732129	&	186.852	&	27.837	&	19.9	&	1051	&	295	&	238	&	29	&	9.80e6	&	0$^{\rm (8b)}$	\\
	&	749238	&	187.376	&	27.617	&	9.3	&	974	&	229	&	161	&	106	&	1.13e7	&	19.1	\\
	&	7673	&	187.995	&	29.714	&	9.7	&	643	&	305	&	-170	&	70	&	1.64e8	&	18	\\
	&	229111	&	188.405	&	29.438	&	18.6	&	1001	&	244	&	188	&	31	&	1.41e7	&	0$^{\rm (8b)}$	\\
\hline																					
4826	&	\textbf{742601}$^{*}$	&	192.401	&	21.918	&	6.4	&	539	&	130	&	131	&	27	&	4.00e6	&	23.6	\\
	&	221120$^{*}$	&	193.924	&	19.21	&	4.8	&	419	&	189	&	11	&	22	&	2.50e7	&	87.3	\\
	&	\textbf{229386}	&	192.941	&	21.735	&	6.9	&	582	&	90	&	174	&	25	&	1.10e7	&	19.2	\\
	&	8030	&	193.62	&	26.303	&	7.8	&	628	&	360	&	220	&	53	&	7.60e6	&	11.1	\\
\hline																																	
\end{tabular}
\begin{tablenotes}
      \small
      \item (1). Satellites that reside in multiple hosts' $2R_{200}$ are assigned to the host with the lower dimensionless separation $\Theta$ as defined in equation \ref{eqn:dimensionless-separation}. (2). ALFALFA source number (AGCNr) of gaseous dwarf galaxies, highlighted in bold are the $R_{200}$ satellites, and with superscripts $^{*}$ are the $V_{\rm esc}$ satellites (see Note 6). (3). The coordinates adopted are $\alpha.100$ HI centroid coordinates (generally agreeing with the optical counterpart coordinates within $0.01{\degr}$). (4). Satellite distances $D_{\alpha}$ from the $\alpha.100$ catalog, which could differ from the host distances to non-negligible values. (5). Projected distance $d_{\rm proj}$ evaluated at the distance of the primary between the satellite-host pair. Satellites with $d_{\rm proj}$ smaller than the host $R_{200}$ (see Table \ref{table:analog_properties}) are the `$R_{200}$' satellites highlighted in bold in column 1. (6). The velocity differences $\delta V = V_{\rm sat} - V_{\rm host}$. If $|\delta V|$ is smaller than the hosts' $V_{\rm esc}$ (Table \ref{table:analog_properties}), the satellite is designated as a `$V_{\rm esc}$' candidate annotated with $^{*}$ in column 1; if $V_{\rm esc} < |\delta V| \leq 300$ km $\rm s^{-1}$, it is a `$V_{300}$' satellite candidate. (7). We scale the ALFALFA $M_{\rm HI}$ values from satellite distances (as listed in this table) to host distances (see Table \ref{table:analog_properties}). (8). The supplementary factor $F_{\rm d_{\rm 3D}}$ (equation \ref{eqn:Fd3d}) defined to compare the relative 3D proximity to the hosts among the satellite candidates. The range $F_{\rm d_{\rm 3D}} = 25 \pm 14\%$ reflects the overall high uncertainties in extragalactic dwarf distances. We flag the four satellite candidates with the lowest 3D proximity factor that will benefit from future deep optical distance follow-up: two are $V_{\rm esc}$ candidates (8a), and two are more likely background objects (8b).
    \end{tablenotes}
\end{threeparttable}
\end{table*} 

\begin{figure*}
    \centering
    \includegraphics[width=0.9\linewidth]{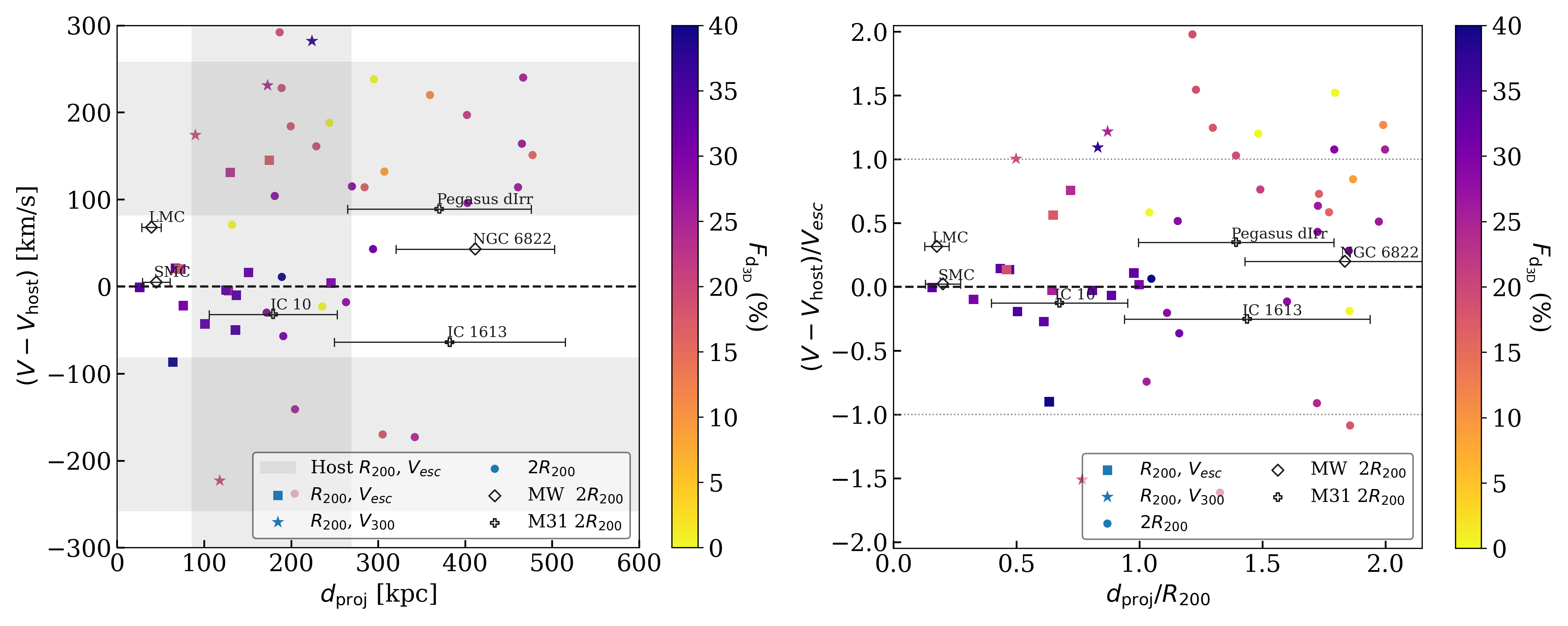} 
    \caption{The velocity separation ($V - V_{\rm host}$) versus projected distance ($d_{\rm proj}$) for the ALFALFA gaseous satellite candidates, colored by $F_{\rm d_{\rm 3D}}$, the relative 3D spatial proximity among the satellite-host pairs (equation \ref{eqn:Fd3d} and Table \ref{table:2R200_satellites}). \textbf{Left panel:} Satellite candidate ($V - V_{\rm host}$) versus $d_{\rm proj}$, where the vertical and horizontal bands show the range of host $R_{200}$ and $V_{\rm esc}$ (Table \ref{table:analog_properties}). The $R_{200}$ satellite candidates under the $V_{\rm esc}$ and $V_{300}$ criteria are marked by squares and stars, respectively (defined in \S \ref{subsec:sat_search}), and those outside of $R_{200}$ while within $2R_{200}$ by filled circles. 
    The empty symbols represent the detectable gaseous satellites of the Milky Way and M31. Their projected distances and $\pm 1\sigma$ uncertainties (from random viewing angles) are obtained from the Monte Carlo mock Local Group observations from ALFALFA host distances; see Section \ref{subsec:lg_dwarf_population}. \textbf{Right panel:} same information, with the velocity separation scaled by the host galaxy's $V_{\rm esc}$ at $R_{200}$, and distance scaled by the host $R_{200}$. The gray dotted lines at $|\delta V|=V_{\rm esc}$ distinguishes, e.g., the `$2R_{200}, V_{\rm esc}$' sample from the `$2R_{200}, V_{300}$' sample.}
    \label{fig:dvel_dproj_scatter}
\end{figure*}

Figure \ref{fig:dvel_dproj_scatter} summarizes the spatial and velocity separations for all the ALFALFA satellite candidates, annotating the Local Group sample (\S\ref{subsec:lg_dwarf_population} and Table \ref{table:LG_2Rvir}) as a reference. The most conservative `$R_{200}, V_{\rm esc}$' satellite selection captures satellite candidates with close separations that resemble the LG $R_{200}$ gaseous satellites (LMC, SMC, and IC 10), and the ALFALFA $2R_{200}$ candidates that satisfy $|\delta V| \leq V_{\rm esc}$ (right panel of Figure \ref{fig:dvel_dproj_scatter}) also have similar separations to the LG $2R_{200}$ satellites (Pegasus dIrr, NGC 6822, IC 1613). The remaining 14/46 satellite candidates have velocity separations $V_{\rm esc} < |\delta V| \leq 300$ km $\rm s^{-1}$ ($V_{300}$), which may indicate higher contamination or complex host environment, primarily affecting NGC 4559 and the NGC 4274 group (9/14 $V_{300}$ objects, see also Figures \ref{fig:analog_mass_stats}, \ref{fig:virgo_field_sky_distribution} and Table \ref{table:2R200_satellites}). 

The coloring of Figure \ref{fig:dvel_dproj_scatter} estimates the ALFALFA satellite candidates' 3D proximity to the host ($F_{\rm d_{\rm 3D}}$, see equation \ref{eqn:Fd3d}). The range of these values ($F_{\rm d_{\rm 3D}} = 25 \pm 14\%$) reflects that the current Mpc-scale line-of-sight distance uncertainties in the data are insufficient for determining satellite 3D association, so we use these as a supplement to the ($d_{\rm proj}, \delta V$)-based satellite selection (\S \ref{subsec:sat_search}). The relative $F_{\rm d_{\rm 3D}}$ values among the candidates independently confirm that the `$R_{200}, V_{\rm esc}$' candidates have the highest confidence in the 3D spatial proximity.

\subsection{Properties of Gaseous Satellite Candidates} \label{subsec:satellite_properties}
\begin{figure}
    \centering
    \includegraphics[width=1.0\linewidth]{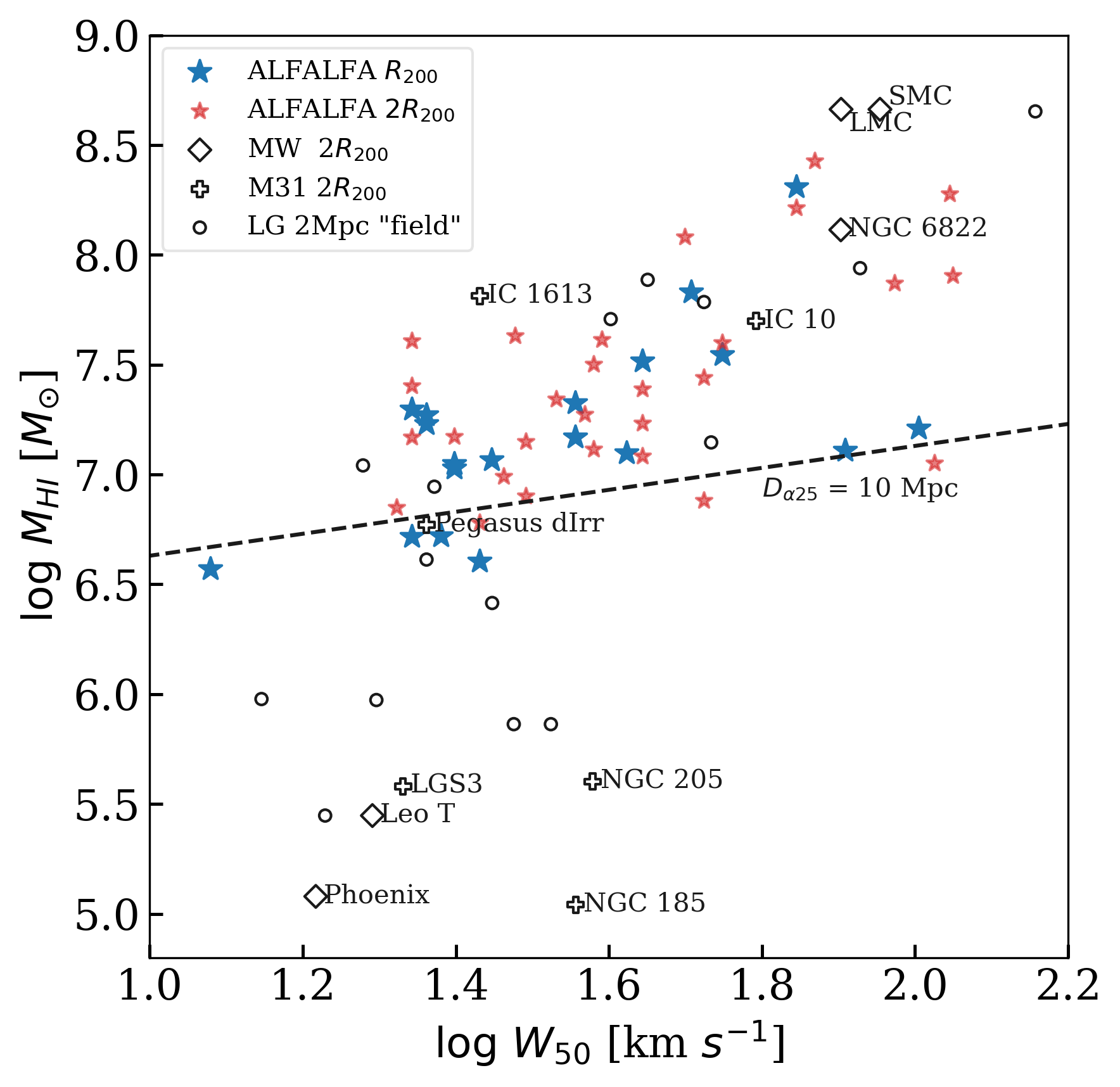} 
    \caption{HI mass and FWHM ($W_{50}$) distribution for the ALFALFA gaseous satellite candidates. 
    The larger blue stars show the ALFALFA $R_{200}$ candidate sample, while the red stars are those outside $R_{200}$ but within $2R_{200}$. We show the Local Group gaseous dwarfs within the 2 Mpc volume from \citet{putman_gas_2021} in empty circles, annotating the $2R_{200}$ satellites of the MW and M31 (\S \ref{subsec:lg_dwarf_population}, Table \ref{table:LG_2Rvir}). The dashed line ($D_{\alpha 25}=10$ Mpc) denotes the ALFALFA $25 \%$ confidence HI detectability limit \citep{haynes_arecibo_2011} evaluated at $10$ Mpc.}
    \label{fig:mhi_w50_scatter_with_sensitivity}
\end{figure}

\begin{figure*}
    \centering
    \includegraphics[width=0.9\textwidth]{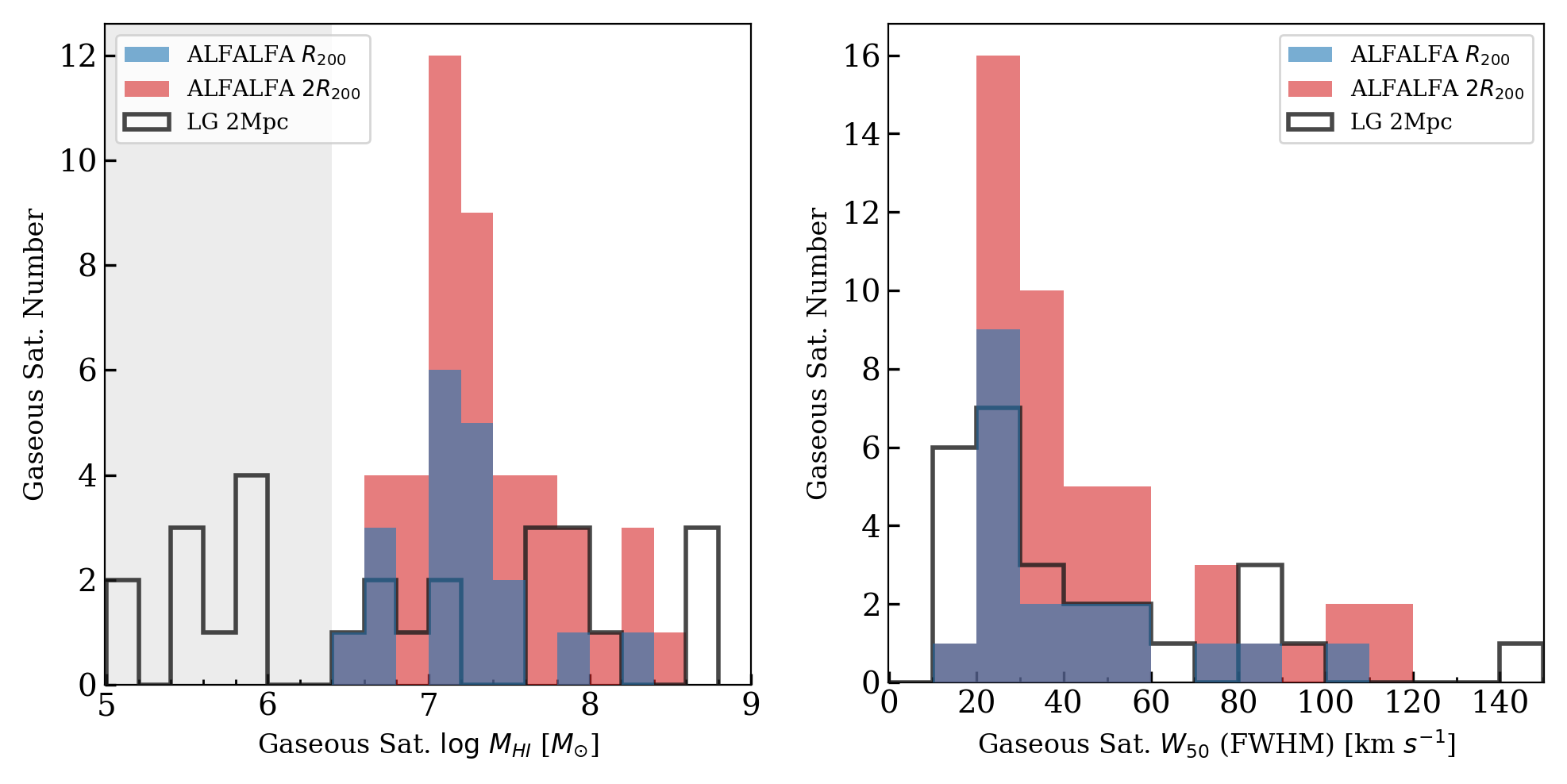}
    \caption{HI mass and FWHM histograms for the gaseous satellite candidates (as in Figure \ref{fig:mhi_w50_scatter_with_sensitivity}). \textbf{Left panel:} the $M_{\rm HI}$ distribution scaled to host distances for ALFALFA $R_{200}$ (blue), $2R_{200}$ (red) satellite candidates. In black is the LG gaseous dwarf population for comparison. The gray shaded region shows the HI mass range where ALFALFA typically would not detect objects at the local volume host distances. \textbf{Right panel:} the HI FWHM ($W_{50}$) for ALFALFA satellite candidates and LG gaseous dwarf galaxies. As discussed in \S\ref{subsec:sat_search}, we excluded massive satellites with $V_{\rm rot}>90$ km $\rm s^{-1}$ for consistency with the LG sample (inclusion of LMC- and SMC-like, exclusion of M33-like dwarf galaxies).}
    \label{fig:mhi_census}
\end{figure*}

The HI properties of the satellite sample are summarized in Figures \ref{fig:mhi_w50_scatter_with_sensitivity} and \ref{fig:mhi_census}. Specifically, we show the HI masses and HI FWHMs ($W_{50}$) for the ALFALFA satellites compared with the full Local Group gaseous dwarf sample within $2$ Mpc. In Figure \ref{fig:mhi_w50_scatter_with_sensitivity}, the Local Group dwarf galaxies within $2$ Mpc but outside of $2R_{200}$ of MW and M31 are here considered the "field" population as opposed to satellite systems. We denote the ALFALFA $25 \%$ detectability level at our host distance cut, $D=10$ Mpc, on the $W_{50}-M_{\rm HI}$ plane. At this distance, ALFALFA is sensitive down to Pegasus dIrr mass $M_{\rm HI} \sim 5.5 \times 10^{6}$ $M_{\odot}$. 
In Figures \ref{fig:mhi_w50_scatter_with_sensitivity}, \ref{fig:mhi_census}, and results hereafter, we have scaled the ALFALFA satellite HI masses to the host distances ($M_{\rm HI} \propto D^{2}$). This only mildly shifts the satellite candidate HI mass distribution to the lower-mass end ($<0.2$ dex) and has no impact on the main results.


Figure \ref{fig:mhi_w50_scatter_with_sensitivity} compares the ALFALFA satellite candidates and the Local Group gaseous dwarfs on an individual galaxy level. It shows that the ALFALFA sample shares typical HI properties with the Local Group gaseous dwarfs above the detectability limit (dashed line in Figure \ref{fig:mhi_w50_scatter_with_sensitivity}), such as the Milky Way and M31 satellites NGC 6822, Pegasus dIrr, IC 10, and IC 1613. However, none of the ALFALFA satellite candidates reach the high HI masses of the Milky Way's Magellanic Clouds ($M_{\rm HI} \approx 10^{8.67}~M_{\odot}$); and of the five most massive candidates with $M_{\rm HI} \gtrsim 10^{8}~ M_{\odot}$, only one falls within the host's virial radius (Figure \ref{fig:mhi_w50_scatter_with_sensitivity}). This agrees with previous findings that the Magellanic system is relatively uncommon for Milky Way-like hosts in a cosmological context (see, e.g., \citealt{liu_how_2011,tollerud_small-scale_2011,robotham_galaxy_2012}). The least HI massive dwarf galaxies, such as Leo T, LGS 3, NGC 185, and NGC 205, are undetectable by ALFALFA at the host distances $D \approx 10$ Mpc (\S \ref{subsec:lg_dwarf_population}). Several ALFALFA satellites with intermediate HI masses ($M_{\rm HI} \sim 10^7$ $M_{\odot}$) and high FWHM's ($W_{50} \sim 100$ km $\rm s^{-1}$) do not seem to have any Local Group counterparts. We checked the optical images and literature inclination angles, if available, of all satellite candidates with a high $W_{50}$ and only included the ones with similar to or less than LMC-like rotation velocities in our sample ($V_{\rm rot} \leq 90$ km $\rm s^{-1}$, see \S\ref{subsec:sat_search}).

Figure \ref{fig:mhi_census} shows the same physical quantities in Figure \ref{fig:mhi_w50_scatter_with_sensitivity} as histograms. In the left panel, we shaded the ALFALFA-undetectable HI mass range in gray, where ALFALFA not detecting objects is an observational sensitivity effect rather than a "missing satellite" problem. The median $\pm 34.1$ percentiles of the two ALFALFA populations are $\log M_{\rm HI} = 7.17 \pm ^{0.35}_{0.45}$ ($R_{200}$, blue) and $\log M_{\rm HI} = 7.23 \pm ^{0.57}_{0.32}$ ($2R_{200}$, red). They are both centrally peaked around the median values. In comparison, the Local Group $2$ Mpc gaseous dwarfs have a relatively flat mass distribution, $\log M_{\rm HI} = 6.86 \pm ^{1.09}_{1.28}$ (LG 2Mpc, black). If we only consider the ALFALFA-detectable part of Local Group galaxies, the distribution becomes $\log M_{\rm HI}(\rm LG_{detectable}) = 7.75 \pm ^{0.70}_{0.91}$, still considerably more spread than ALFALFA. We note that these are small number statistics and likely skewed by the Milky Way's Magellanic Clouds, which are less common for the ALFALFA spirals. The right panel of Figure \ref{fig:mhi_census} shows that the HI velocity width distributions are similar among all dwarf samples\footnote{Andromeda's massive satellite, NGC 3109, has a high $W_{50}$ of $144$ km $\rm s^{-1}$ but is almost edge-on \citep{barnes_neutral_2001,carignan_kat-7_2013}. With the inclination correction, it satisfies our $V_{\rm rot} \leq 90$ km $\rm s^{-1}$ criterion.}, peaking at $20 \leq W_{50} \leq 30$ km $\rm s^{-1}$.

\begin{figure}
    \centering
    \includegraphics[width=1.0\linewidth]{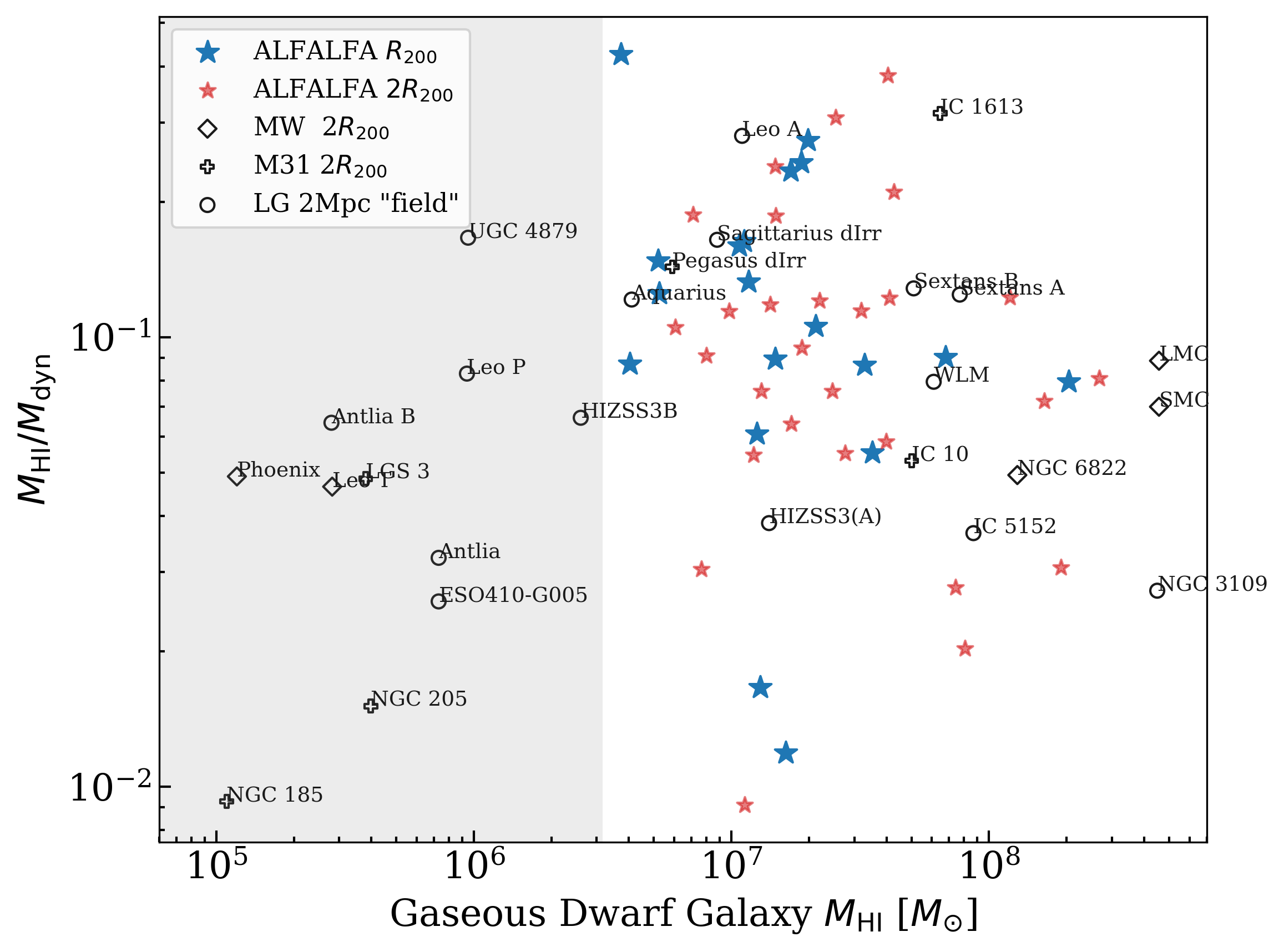}
    \caption{The estimated $M_{\rm HI}/M_{\rm dyn}$ ratio versus HI mass for the gas-containing dwarf galaxies around the ALFALFA local universe spiral hosts in comparison with the LG gaseous population. 
    The ALFALFA $R_{200}$, $2R_{200}$ satellite candidates and the LG gaseous dwarf galaxies are denoted in the same style as in Figure \ref{fig:mhi_w50_scatter_with_sensitivity}. The region undetectable by ALFALFA due to the HI sensitivity limit is shaded as in Figure \ref{fig:mhi_census}. The dynamical (`total') mass $M_{\rm dyn}$ estimates are consistently derived from the gaseous dwarf galaxies' HI observables.}
    \label{fig:dwarf_galaxy_HI_depletion_lg_comparison}
\end{figure}

We further use the HI properties to estimate the total dynamical masses ($M_{\rm dyn}$) of the gaseous dwarf sample (Figure \ref{fig:dwarf_galaxy_HI_depletion_lg_comparison}). The dynamical mass estimates are based on a variation of the optical method commonly adopted for Local Group studies \citep{mcconnachie_observed_2012,putman_gas_2021}: $M_{\rm dyn}(r_{\rm half})=\mu \cdot r_{\rm half} \cdot \sigma^{2}$ \citep{walker_universal_2009}. For $\sigma$, we use the gas velocity dispersion $\sigma_{\rm gas}=W_{50}/2.355$ (stellar velocity dispersion not readily available for ALFALFA dwarf galaxies). And in place of the optical half-light radius $r_{\rm half}$, we use the gaseous disk radius ($r_{\rm HI}$) from the observed HI mass-radius relation, $\log (M_{\rm HI})=1.86 \log (D_{\rm HI}) + 6.6$ \citep{swaters_westerbork_2002}.  The ALFALFA satellites are not resolved, and we use this relation for the Local Group satellites for consistency.  The dynamical mass can be expressed as $M_{\rm dyn}/M_{\odot}=698 \cdot (r_{\rm HI}/ \rm pc) \cdot (\sigma_{\rm gas} /(\rm km \cdot \rm s^{-1}))^{2}$. We note that the $M_{\rm dyn}$ expression is only an indicator of the dwarf galaxy's total mass within the gaseous radius and is not a direct measurement of the dark matter halo mass. Obtained directly from HI observables, it offers a self-consistent comparison amongst gaseous dwarf populations regarding the gas-to-total mass fraction ($M_{\rm HI}/M_{\rm dyn}$).

Figure \ref{fig:dwarf_galaxy_HI_depletion_lg_comparison} shows the resulting $M_{\rm HI}/M_{\rm dyn}$ ratios for the gaseous dwarfs. The ratio can differ significantly, ranging from $>10\%$ for the star-forming dIrr's, such as IC 1613 and Leo A, to $\lesssim 1\%$ for the dE's that retain very small amounts of gas content and are not star-forming, such as Andromeda's close satellites NGC 185 and NGC 205. Despite the large scatter among individual galaxies, there is an agreement in the average $M_{\rm HI}/M_{\rm dyn} \approx 12\%$ between our ALFALFA sample and the ALFALFA-detectable Local Group sample, indicating a similar average gas abundance between the two. In contrast, the low-mass LG dwarf population undetectable by ALFALFA (shaded in Figure \ref{fig:dwarf_galaxy_HI_depletion_lg_comparison}) has a distinctively lower average ratio, $M_{\rm HI}/M_{\rm dyn} \approx 5.5\%$. We note that this method has potential caveats. For example, of the two ALFALFA $R_{200}$ objects with the lowest $M_{\rm HI}/M_{\rm dyn}$ (two blue stars at the bottom of Figure \ref{fig:dwarf_galaxy_HI_depletion_lg_comparison}), one is a particularly close dwarf satellite with low HI, but the other is an edge-on spiral, and not applying the inclination correction gives it an unusually high $M_{\rm dyn}$. The average gas-to-stellar mass ratio ($M_{\rm HI}/M_{*}$) of the ALFALFA gaseous satellites, where available, also agrees with the detectable LG satellites, but we leave detailed stellar mass analyses to future studies because the optical data are currently incomplete for the ALFALFA dwarfs (missing $\approx 40\%$, see Figure \ref{fig:alfa_elves_saga_luminosity_comparison}).

\subsection{Number of Gas-containing Satellites Per Host} \label{subsec:satellite_number}
This section summarizes the number of HI-containing satellites ($N_{\rm sat}$) versus the host halo mass (see Table \ref{table:analog_properties}) for the ALFALFA primaries in comparison with MW and M31 observations (\S \ref{subsec:lg_dwarf_population}) and the Auriga cosmological zoom simulations of Milky Way analogs \citep{grand_auriga_2017-1,simpson_quenching_2018}. We chose the Auriga simulation suite because of its overall consistency with LG observations and the availability of satellite rotation velocity data ($V_{\rm rot}$). Recent work has found Auriga to be broadly consistent with other high-resolution cosmological simulations of Milky Way analogs, such as FIRE-2 \citep{hopkins_fire-2_2018,samuel_extinguishing_2022} and TNG50 \citep{nelson_first_2019,pillepich_first_2019,engler_abundance_2021-1}. 
The following section (\S \ref{sec:elves_saga_discussion}) is dedicated to a comprehensive comparison with recent deep optical surveys of satellite systems around MW-analogs, the ELVES survey \citep{carlsten_exploration_2022-1} and the SAGA survey \citep{geha_saga_2017,mao_saga_2021}.

\begin{figure*}
    \centering
    \includegraphics[width=0.9\linewidth]{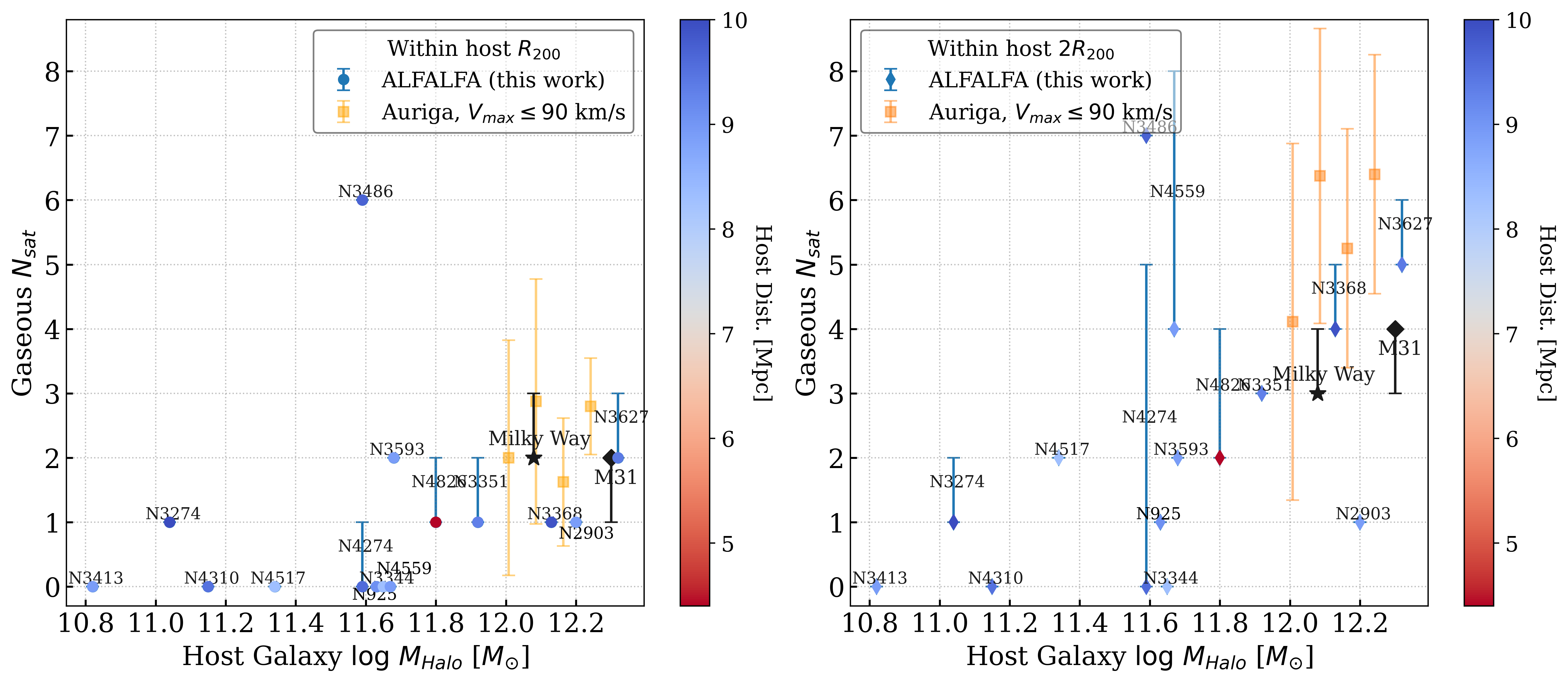}
    \caption{Number of gaseous satellites per spiral primary within the host's projected $R_{200}$ (left panel) and $2R_{200}$ (right panel) versus host halo mass. For our ALFALFA spiral hosts, the data points (color-coded by distances, see Table \ref{table:analog_properties}) show the number of the $V_{\rm esc}$ satellites, and the upper error bars show $V_{300}$ (defined in \S \ref{subsec:sat_search}).
    The summary statistics of the Milky Way and M31's detectable gaseous satellites are denoted in black (data points: median; $\pm~34.1$ percentiles: error bars), which are obtained from the Monte Carlo mock LG observations (\S \ref{subsec:lg_dwarf_population}). Orange squares are the binned number of HI-rich satellites with rotation velocities $V_{\rm rot} \leq 90$ km $\rm s^{-1}$ per Auriga MW analog (Table 1 in \citealt{simpson_quenching_2018} and C. Simpson priv. comm.).}
    \label{fig:count_comprehensive}
\end{figure*}

Figure \ref{fig:count_comprehensive} shows a consistency in the gaseous $N_{\rm sat}$ among the ALFALFA local volume spirals and the MW and M31. Within the virial radii (left panel), the typical gaseous satellite number is $N_{\rm sat}=0-3$ (for 14/15 ALFALFA hosts), which agrees with the detectable $N_{\rm sat}$ for the MW ($N=2_{-0}^{+1}$: LMC, SMC, and potentially NGC 6822) and M31 ($N=2_{-1}^{+0}$: IC 10 and either IC 1613 or Pegasus dIrr, depending on the viewing angle; see Section \ref{subsec:lg_dwarf_population}). The ALFALFA primary NGC 3486 appears to be an outlier in $R_{200}$ with a high $N_{\rm sat}=6$, but we note that four of the satellite candidates are less likely dwarf galaxies (see \S\ref{subsec:NGC3486}). Extending out to $2R_{200}$ (right panel), 12/15 of the ALFALFA hosts contain $0-5$ satellite candidates with a median value of $2$, also agreeing with the MW and M31's detectable numbers ($N = 3_{-0}^{+1}$ and $N = 4_{-1}^{+0}$ respectively; Section \ref{subsec:lg_dwarf_population}). Other than NGC 3486, two more ALFALFA hosts, NGC 4559 and NGC 3627 (M66, group primary of Leo Triplet), appear to be physically explainable outliers with higher $N_{\rm sat}$ values within $2R_{200}$.  Both of these hosts have distinctively complicated group environments and are likely contaminated at these larger radii. These types of outliers have been found in other spiral studies \citep{spencer_survey_2014,smercina_saga_2020,mao_saga_2021,carlsten_exploration_2022-1}. In the following, we focus on global results and leave the discussion of individual host environments to Appendix \S \ref{sec:results_individual_hosts}.

Within our ALFALFA host sample, there is a moderate trend of increasing $N_{\rm sat}$ with higher halo mass for both the $R_{200}$ and $2R_{200}$ samples (Figure \ref{fig:count_comprehensive}). The trend becomes more significant if we take out the physically explainable potential outlier host NGC 3486 or only consider the $V_{\rm esc}$ (higher confidence of physical association) satellites. We also notice that when scaling $N_{\rm sat}$ by the host galaxies' projected sky area, the increasing satellite with host mass trend flattens significantly. Future data and a larger sample size are needed for further quantification of the trend and the potential contamination effects.

When comparing with Auriga simulations, we cut the Auriga gaseous satellites by rotation velocity for consistency with our ALFALFA selection (requiring $V_{\rm rot} \leq 90$ km $\rm s^{-1}$; \S \ref{subsec:sat_search}), 
which reduced the $N_{\rm sat}$ per Auriga primary by $0.5$ (for $R_{200}$) and $1$ (for $2R_{200}$). No completeness/detectability correction is necessary here because the HI masses of the included Auriga satellites are almost always detectable for ALFALFA sensitivity (figure 3 of \citealt{simpson_quenching_2018}).  We note that we use the satellites within the actual volumes for the simulations, while for ALFALFA, it is the satellites within the projected volumes. Figure \ref{fig:count_comprehensive} shows that the Auriga primaries have $N_{\rm sat}(R_{200}) \sim 0-5$ per host with a median of $2$ within the virial radii, which agrees with our ALFALFA results and with LG observations. Extending out to $2R_{200}$, the Auriga numbers increase to $N_{\rm sat}(2R_{200}) \sim 3-9$ for 28 of the 30 primaries with a median of $5.5$, showing an excess compared to the 3D $2R_{200}$ satellites of the Milky Way and Andromeda. In both simulations and observations, the average gaseous satellite number per unit volume decreases when extending to $2R_{200}$. While this is expected because, at increasing distances to the central halo, $\Lambda$CDM predicts a lower number density of subhalos (e.g., \citealt{diemand_velocity_2004,springel_aquarius_2008,guo_satellite_2012}); it may be unexpected given major gas stripping mechanisms are less effective at larger distances (\S \ref{sec:intro}).

\section{Direct Comparison to ELVES and SAGA} \label{sec:elves_saga_discussion}

Beyond the Milky Way and M31, we can now compare our results to satellites around other spiral galaxies with recent deep optical surveys. In particular, we focus on the results of the recently completed ELVES survey within the local volume $0<D<12$ Mpc \citep{carlsten_exploration_2022-1}, and the ongoing SAGA survey within a slightly higher redshift range placing the Milky Way analogs at $20<D<40$ Mpc \citep{geha_saga_2017,mao_saga_2021}.

\subsection{Selecting Satellite Analogs}
Built upon previous wide-field optical imaging and dwarf galaxy distance measurements \citep{carlsten_wide-field_2020,carlsten_radial_2020,carlsten_luminosity_2021}, the ELVES survey catalogs a distance-confirmed list of 251 dwarf satellites around 25 cataloged host galaxies in the local volume ($D<12$ Mpc). 
Their Local Volume host sample is selected by a relatively simple cut in luminosity $M_{K_{s}} < -22.1$ $\rm mag$ that includes a few lenticular/elliptical hosts. We take out these early-type hosts when comparing with our ALFALFA spiral host sample. Within their spiral host galaxies, five (NGC 2903, NGC 3344, NGC 3627, NGC 4517, and NGC 4826) are direct overlaps with our host sample. 
ELVES visually classifies the dwarf sample to have early- or late-type morphology. We adopt their late-type classification as an indicator of the presence of gas content for comparisons hereafter.

The SAGA survey constructs a growing sample size of MW-like host systems based on a combination of K-band luminosity and local environment criteria \citep{geha_saga_2017,mao_saga_2021}, selecting a host list most resembling the MW in terms of mass and environment compared to ELVES and our ALFALFA sample. Its satellite completeness level of $M_{r} < -12.3$ $\rm mag$ is about 3 $\rm mag$ brighter than the more local ELVES sample that is complete to $M_{V} \sim -9$. We classify the SAGA satellites as gaseous based on the SAGA team's H-$\alpha$ line criterion, for galaxies with H-$\alpha$ emission are usually observable in HI and vice versa \citep{meurer_survey_2006, van_sistine_alfalfa_2016}.

\begin{figure}
    \centering
    \includegraphics[width=1.0\linewidth]{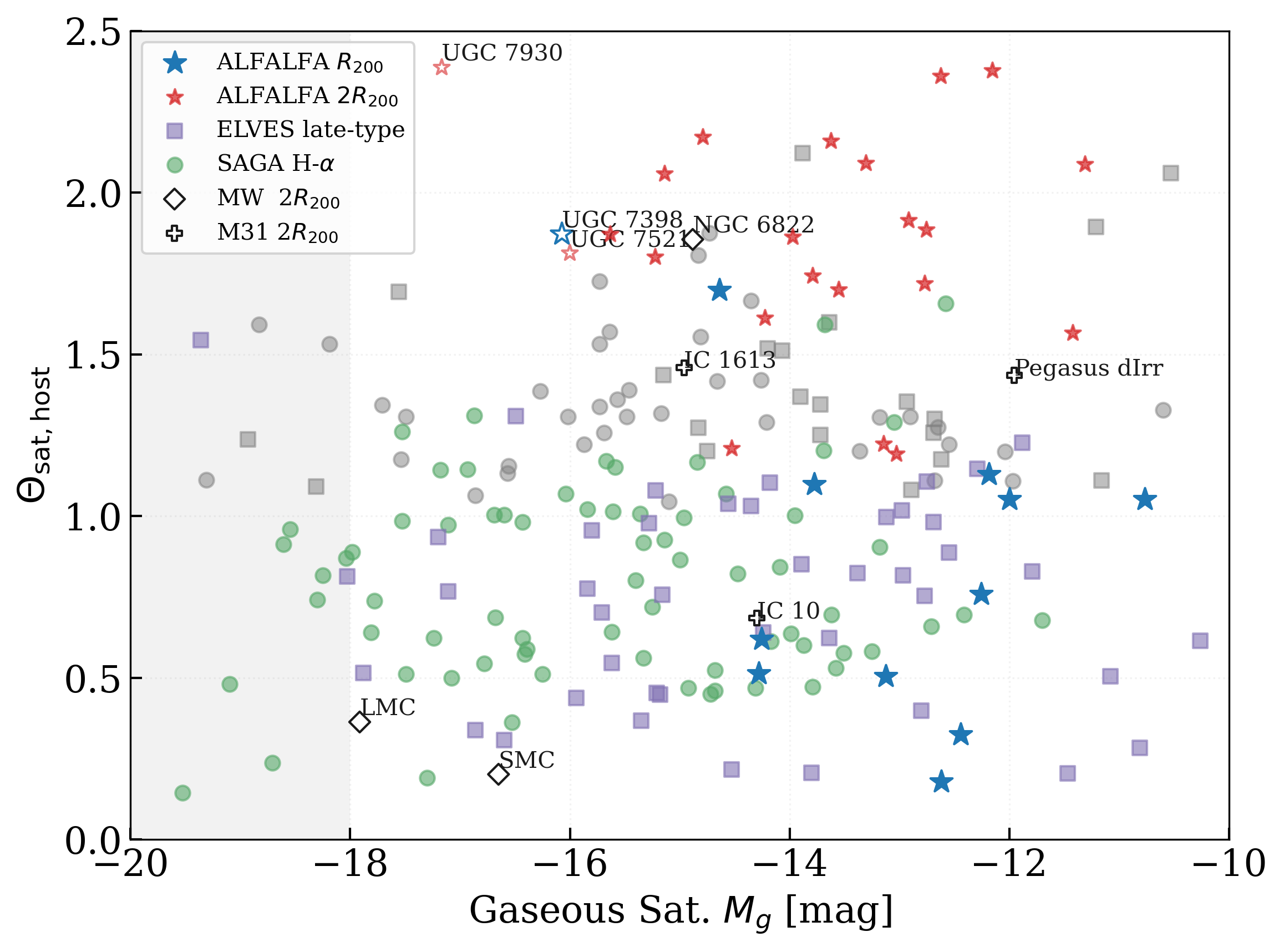}
    \caption{Satellite-host pair kinematic separation $\Theta$ (equation \ref{eqn:dimensionless-separation}) versus g-band absolute magnitudes $M_{g}$ for the gaseous/star-forming samples. The ALFALFA luminosity data from ALFALFA-SDSS \citep{durbala_alfalfa-sdss_2020-1} are $58 \%$ complete for the $R_{200}$ sample and $67 \%$ complete for $2R_{200}$ (blue and red filled stars, respectively). Empty stars show the massive ALFALFA candidates excluded from our satellite sample due to $V_{\rm rot} > 90$ km $\rm s^{-1}$. 
    The MW and M31's detectable gaseous satellites are shown in the same style as Figure \ref{fig:dvel_dproj_scatter}.
    The gaseous/star-forming population of ELVES and SAGA satellites within host $R_{200}$ are denoted by magenta squares and green circles, respectively, and the ones beyond $R_{200}$ in gray. Those ELVES and SAGA star-forming satellites beyond the host's $R_{200}$ or brighter than the LMC ($M_{g} \leq -18$, shaded) are not included in other plots.}
    \label{fig:alfa_elves_saga_luminosity_comparison}
\end{figure}

Figure \ref{fig:alfa_elves_saga_luminosity_comparison} summarizes the gaseous satellite luminosities and the host-satellite separations across the different studies. To compare with our ALFALFA dwarf satellites that are HI gas-containing and have up to LMC-like total masses (\S \ref{subsec:sat_search}), we apply three filters on the ELVES and SAGA dwarfs: (i). distinguish whether they reside within the host's virial radius; (ii). select the late-type/star-forming satellites from ELVES and SAGA, respectively, to represent the gaseous population; and (iii). apply a g-band absolute luminosity cut at the LMC value to exclude the brighter-than-LMC satellites. The luminosity cut in (iii) is used because of the lack of HI velocity width or rotation velocity information within these optical surveys. We obtain the g-band luminosities of the MW and M31's $2R_{200}$ satellites from \citet{mateo_dwarf_1998} and \citet{mcconnachie_observed_2012}, and the LMC is the brightest at $M_{g}\rm (LMC) \approx -17.9$. We hence place the cut of $M_{g}>-18$ for ELVES and SAGA satellites to select dwarf galaxies up to LMC-like luminosity. Globally, 202/251 of the ELVES satellites reside within $R_{200}$, where 62 out of the 202 are classified as late-type, and 54 are late-type dwarfs dimmer than the LMC; 83/127 of the SAGA satellites reside within $R_{200}$, 71 of which are star-forming, and 63 are star-forming and dimmer than the LMC. The SAGA hosts are largely Milky Way mass, and many star-forming satellites are cut because they lie between 224 (our adopted MW virial radius) and 300 kpc.

Figure \ref{fig:alfa_elves_saga_luminosity_comparison} shows that the ALFALFA gaseous sample is somewhat less luminous compared to the inferred gaseous populations in ELVES and SAGA, with the ALFALFA $2R_{200}$ sample showing slightly more scatter than the $R_{200}$ sample. Between the two optical surveys, ELVES (completeness limit $M_{V} \sim -9$) captures more low-luminosity gaseous satellites than SAGA (completeness limit $M_{r} \sim -12.3$). Following methods in \S \ref{sec:methods_spiral_host_definition}, we derive $M_{\rm Halo}$, $R_{200}$, and $V_{\rm esc}$ at $R_{200}$ for ELVES and SAGA primary galaxies based on their stellar masses $M_{*}$ (Table 1 in \citealt{carlsten_exploration_2022-1}, and M. Geha priv. comm.). The dimensionless $\Theta_{\rm sat,host}$ value (equation \ref{eqn:dimensionless-separation}) demonstrates the separation between host-satellite pairs that is normalized by host galaxy size. Our $R_{200}$ satellites share a similar $\Theta$ separation range with the $R_{200}$ samples of ELVES and SAGA, which both focus on the projected virial areas of the hosts.  

\subsection{Number of Star-Forming/Gaseous Satellites}\label{subsec:sf_sats_elves_saga}
\begin{figure}
    \centering
    \includegraphics[width=1.0\linewidth]{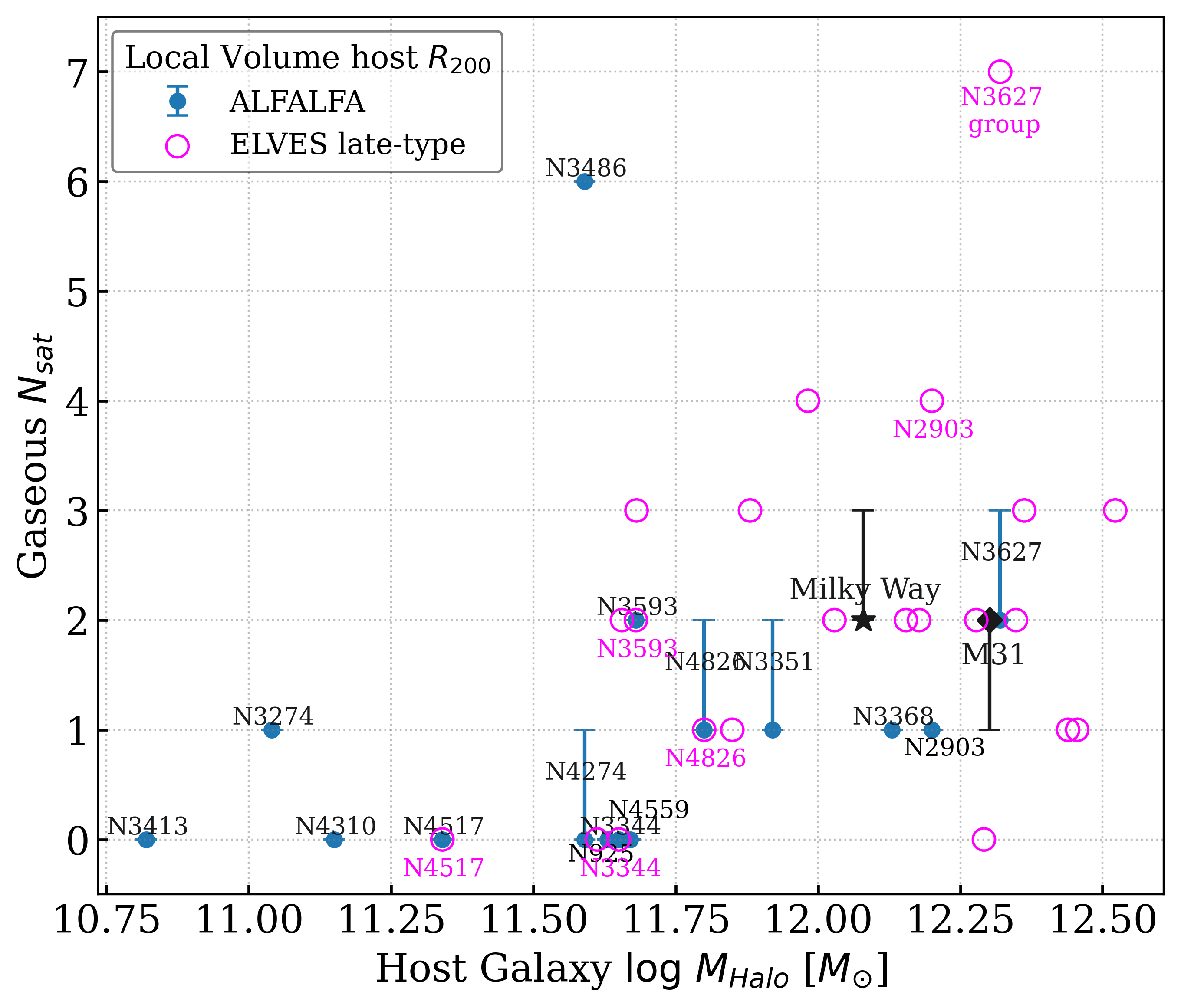}
    \caption{Gaseous satellite number $N_{\rm sat}$ within $R_{200}$ versus host halo mass for the Local Volume spiral hosts. The $N_{\rm sat}$ points for ALFALFA, MW, and M31 are in a consistent style with Figure \ref{fig:count_comprehensive}, and the number of gaseous satellites with $M_{g} > -18$ for the ELVES late-type hosts are shown in magenta empty circles. For the ELVES hosts within the ALFALFA primary list, we take the ALFALFA stellar masses consistent with Table \ref{table:analog_properties}, infer the halo masses, and annotate the NGC numbers in magenta, while for other ELVES hosts, we take the ELVES stellar masses and infer the halo masses. NGC 3627 is the primary of a small group (see \S \ref{subsec:M66Group}), and here we separate the ELVES satellites between NGC 3627 and NGC 3593 based on our $\Theta$ proximity criterion (equation \ref{eqn:dimensionless-separation}).}
    \label{fig:alfa_elves_host_direct_comparison}
\end{figure}

 This section compares the gaseous satellite numbers per host in ELVES and SAGA with the ALFALFA sample. As stated above, the ELVES and SAGA satellites are inferred to contain HI gas if they are late-type or detected in H$\alpha$ emission, respectively. ELVES and this ALFALFA work both target spiral galaxies in the local volume, which offers a complementary view of the local picture of gaseous satellites. Figure \ref{fig:alfa_elves_host_direct_comparison} shows the host-to-host scatter of the ALFALFA and ELVES gaseous satellite numbers. We only include late-type primary hosts with confirmed satellites from the ELVES survey and apply the $M_{g} > -18$ satellite cut as discussed above. We see an overall agreement in typical gaseous $N_{\rm sat}$ between ALFALFA ($0-3$ within $R_{200}$) and ELVES ($0-4$ within $R_{200}$; excluding NGC 3627, see below). The MW and M31's gaseous satellite numbers also appear typical among these Local Volume spirals. Compared to the mild mass trend seen in ALFALFA hosts, ELVES hosts show a more significant increase in gaseous satellite abundance with increasing host mass. This increasing $N_{\rm sat}$ with host mass trend is also noted in their complete satellite (late- and early-type dwarfs at all magnitudes) sample \citep{carlsten_exploration_2022-1}. 

We annotated the overlapping spiral hosts between ELVES and ALFALFA in Figure \ref{fig:alfa_elves_host_direct_comparison} for an individual galaxy comparison. Both surveys found $N_{\rm sat}=0$ for the two lowest mass hosts, NGC 4517 and NGC 3344. For NGC 4826, one of our two gaseous satellites is the single late-type satellite ELVES found. For NGC 2903, we only capture the most massive of the four ELVES satellites. The remaining three have stellar masses of $\log M_{*}/M_{\odot}=5.64$, $5.98$, and $6.78$, respectively. Assuming $M_{\rm HI}/M_{*} \approx 1$ places all three below or close to the ALFALFA sensitivity limit (Figure \ref{fig:survey_sensitivity_with_count}) at the host distance $D \approx 9$ Mpc. NGC 3627 is a small group primary whose $R_{200}$ halo overlaps with NGC 3593 (Figure \ref{fig:virgo_field_sky_distribution}). ELVES assigns all the group satellites to the primary, but for consistency, we separate its satellites from NGC 3593 using the $\Theta$ proximity selection (equation \ref{eqn:dimensionless-separation}). After re-assigning the satellites to the closer host, we confirm both ALFALFA NGC 3593 satellites and all three ALFALFA NGC 3627 satellites are found in ELVES. However, as can be seen from Figure \ref{fig:alfa_elves_host_direct_comparison}, we are missing four of the ELVES-classified late-type satellites of NGC 3627. Three of these missing satellites are somewhat less massive with stellar masses of $\log M_{*}/M_{\odot}=7.26$, $6.36$, and $6.79$, and are likely near or below the ALFALFA limit; the fourth satellite is IC 2787, which is sometimes classified as an early-type dwarf \citep{grossi_hi_2009}. Of the overlapping hosts, 7/8 ALFALFA satellites within $R_{200}$ are confirmed by ELVES; the remaining 1 is rejected by ELVES based on an updated distance measurement (see \S \ref{appendix:Fd3d}, \S \ref{subsec:ngc4826}; \citealt{carlsten_exploration_2022-1}).

\begin{figure}
    \centering
    \includegraphics[width=1.0\linewidth]{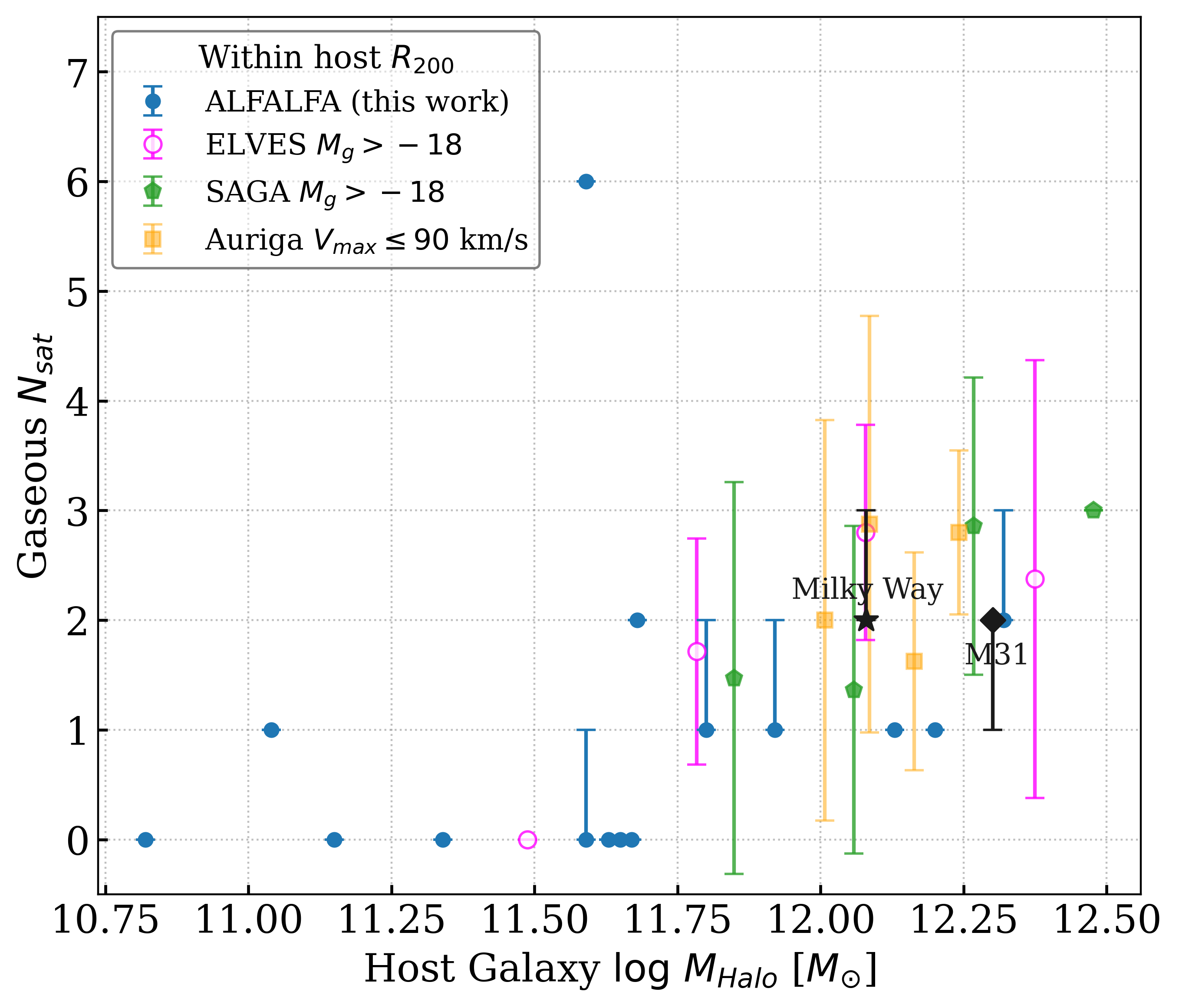}
    \caption{Gaseous satellite abundance $N_{\rm sat}$ within $R_{200}$ versus host halo mass across different spiral systems. The ALFALFA spiral hosts, the Milky Way, and M31 are in the same style as Figures \ref{fig:count_comprehensive} and \ref{fig:alfa_elves_host_direct_comparison}.
     For the deep optical surveys ELVES and SAGA and the Auriga simulation suite, we bin the host galaxy data ($\approx 30$ hosts for each study) by their halo masses.}
    \label{fig:Mg_cut_satellite_count_Rvir_comprehensive}
\end{figure}

Figure \ref{fig:Mg_cut_satellite_count_Rvir_comprehensive} summarizes the gaseous satellite number within the host galaxy's virial radius across different studies of spiral systems. There is an overall agreement of $0-4$ gaseous satellites per primary system and a mild trend of increasing satellite abundance with higher halo mass. Our sample extends to lower host galaxy mass ranges than other studies and finds that relatively isolated low-mass spiral systems in the Local Volume tend not to have any gas-containing dwarf satellites. The Milky Way analogs -- in the Local Volume, at slightly higher redshift, and in cosmological hydrodynamical simulations -- tend to have $1-3$ detectable gaseous satellites on average, which is in good agreement with Local Group results. We performed the same analysis using a uniform spatial selection of $300$ kpc (commonly adopted as the virial radius of Milky Way analogs; \S \ref{subsec:sat_search}) while excluding the low-mass ALFALFA hosts that are not Milky Way-like (those with $\log M_{*}<10$; Table \ref{table:analog_properties}). Within 300 kpc, there is an agreeing gaseous satellite abundance $N_{\rm sat}$ of $1-5$ despite a large host-to-host scatter.

\subsection{Assessing the Quenched Fraction}  \label{subsec:quenched_fraction}
Aside from the satellites containing gas, we expect a large number of satellites without gas (i.e., quenched) for spiral galaxies like the MW and M31 \citep{grcevich_h_2009,2015ApJ...808L..27W,simon_faintest_2019-1,putman_gas_2021}. 
Beyond the Local Group, different surveys find a variety of quenched fractions. The Auriga simulations' average of $\sim 73\%$ quenched fraction within $300$ kpc (their adopted $R_{\rm vir}$, see Table 1 in \citealt{simpson_quenching_2018}) agrees with the Local Group. The ELVES survey has an agreeing quenched fraction ($\sim 50-80\%$), which also shows a mild increasing trend with host luminosity \citep{carlsten_exploration_2022-1}. 
The SAGA survey finds a distinctively lower overall quenched fraction ($\sim 15\%$), but it also shows consistent gaseous satellite numbers in our selected luminosity range ($M_{g}$ dimmer than $-18$) within $R_{200}$ and $300$ kpc (Figure \ref{fig:Mg_cut_satellite_count_Rvir_comprehensive} and \S \ref{sec:elves_saga_discussion}). The SAGA survey targets a larger distance range, where detecting the (predominantly quenched) low surface brightness population is more challenging. For the more massive satellites ($\log M_{*} \in [6.75, 9.5]$), the SAGA quenched fraction $\sim 30\%$ is more comparable to ELVES at similar host luminosities \citep{mao_saga_2021,carlsten_exploration_2022-1}.

 The quenched fraction for the ALFALFA hosts that overlap with ELVES can be directly noted within $R_{200}$ (Figure \ref{fig:alfa_elves_host_direct_comparison}). The two least massive overlapping hosts, NGC 4517 and NGC 3344, have no gaseous satellites within $R_{200}$ in both ALFALFA and ELVES and have $1$ and $3$ ELVES-identified quenched satellites. For NGC 4826, ELVES found 3 quenched satellites, giving a $75 \%$ quenched fraction consistent with the MW or M31 values. For the isolated spiral NGC 2903, where ELVES identified three more low-mass gaseous satellites than ALFALFA, two quenched satellites are found, which gives a $33 \%$ quenched fraction.  We excluded NGC 3627 (Leo Triplet primary) and NGC 3593 here because hosts in groups are especially challenging due to different studies' choices of satellite assignment (see \S \ref{subsec:sf_sats_elves_saga}). ELVES found 8 quenched satellites in total for this system.

\begin{figure}
    \centering
    \includegraphics[width=1.0\linewidth]{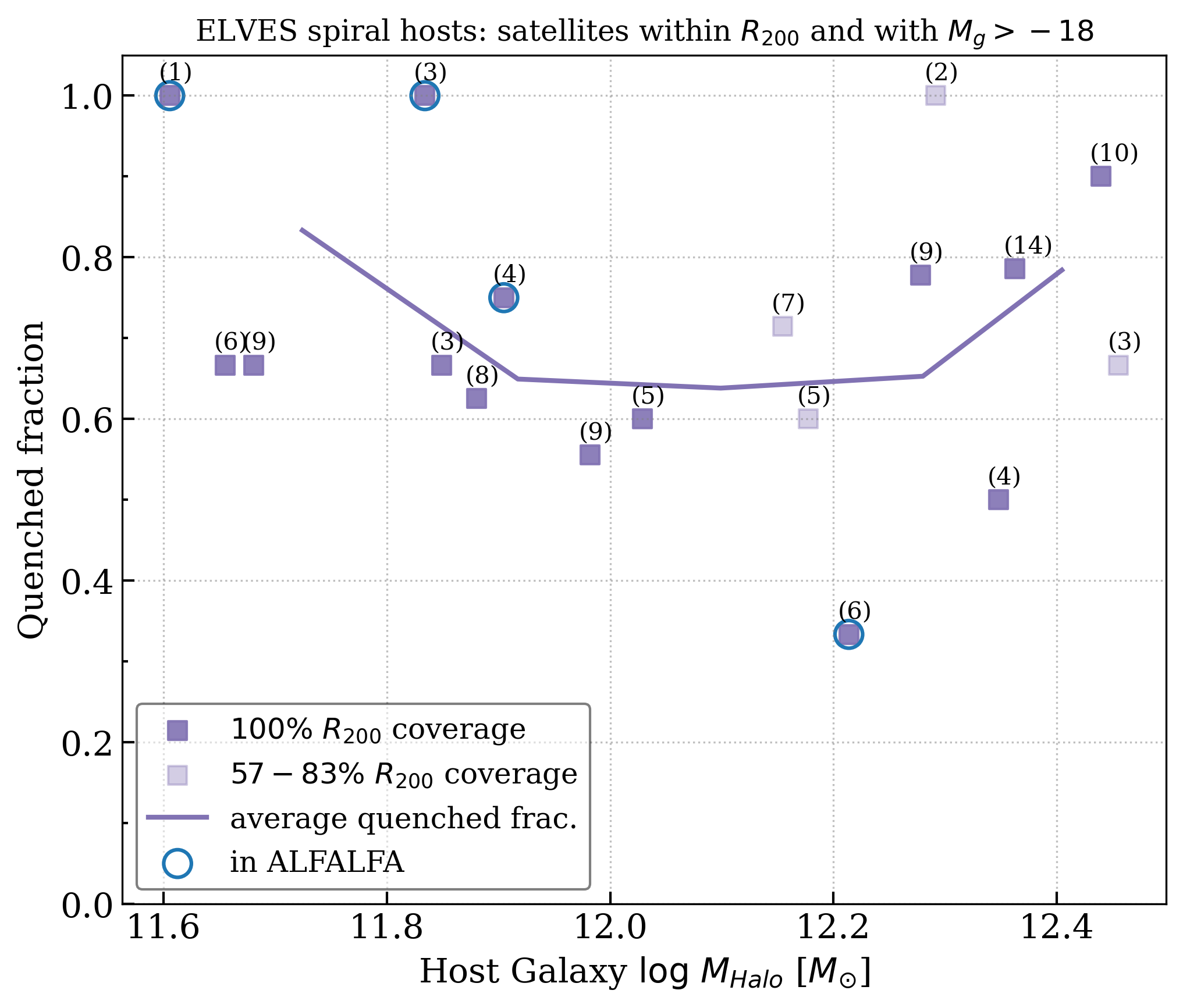}
    \caption{Quenched fractions of the local volume spiral galaxies that have $>50\%$ $R_{200}$ coverage in the ELVES survey \citep{carlsten_exploration_2022-1}, highlighting those in our ALFALFA sample with the blue circles. We mark the few hosts with a partial $R_{200}$ coverage in lighter purple. Following our spatial and luminosity selections (\S \ref{sec:elves_saga_discussion}), we only include satellites within the host's projected virial radius $R_{200}$ and dimmer than $M_{g}=-18$ (see Figure \ref{fig:alfa_elves_saga_luminosity_comparison}). Satellite numbers per host under this criterion are annotated in parentheses.}
    \label{fig:quenched_fraction_spatial_and_luminosity_cut}
\end{figure}

Generally, future deep optical data are needed to assess the gasless satellite population and the quenched fraction, but predictions can be made using the existing optical surveys. The ELVES full sample, within 300 kpc or occasionally smaller spatial area, shows a mild increasing trend of quenched fraction with host K-band luminosity \citep{carlsten_exploration_2022-1}. This trend, although unaffected by the satellite luminosity cut ($M_{g}>-18$), flattens out under the projected virial radius cut. In particular, the least massive hosts increase from $40-50\%$ to $80-100\%$ quenched when scaling from 300 kpc to $R_{200}$; the Milky Way-like hosts -- with their $R_{200}$ closer to 300 kpc -- do not shift as much. The resulting ELVES average quenched fraction, $70.3 \pm 15.2 \%$, is almost independent of the host halo mass (Figure \ref{fig:quenched_fraction_spatial_and_luminosity_cut}). The average fraction predicts that an ALFALFA host with $1-3$ gaseous satellites could have, on average, $2-7$ quenched satellites at the ELVES depth limit. But scatter among individual host systems is significant at these small sample sizes as seen in the ALFALFA direct overlaps (Figure \ref{fig:quenched_fraction_spatial_and_luminosity_cut}), which points to the importance of understanding local environmental effects in addition to building a larger sample.

\section{Summary and Future Work}\label{sec:summary}

Our goal in this project was to compare the ALFALFA local spiral galaxies with the Milky Way and M31 in terms of their abundance of gaseous dwarf satellites. Our results indicate that the local universe spiral galaxies of MW mass and lower (15 total) tend to have very few gaseous satellites. 
We find $0-3$ gaseous satellite candidates within the spiral hosts' virial radii ($R_{200}$) and about $0-5$ within $2R_{200}$ at our completeness limit. These numbers are consistent with the detectable numbers for the MW and M31 (Figure \ref{fig:count_comprehensive}).

Beyond the MW and M31, we further compared our results with two recent deep optical surveys: ELVES \citep{carlsten_exploration_2022-1} in the Local Volume, and SAGA \citep{geha_saga_2017,mao_saga_2021} at a slightly higher redshift; and with the Auriga cosmological zoom simulations of Milky Way-analogs \citep{grand_auriga_2017-1,simpson_quenching_2018}. For Auriga, we applied the same satellite rotation velocity cut $V_{\rm rot}\leq 90$ km $\rm s^{-1}$ as ALFALFA to select dwarf satellites with up to LMC-like masses (\S \ref{subsec:sat_search}). For ELVES and SAGA, we instead applied a dwarf luminosity cut $M_{g}>-18$ on the gaseous satellites (Figure \ref{fig:alfa_elves_saga_luminosity_comparison}). Within the projected virial radii of the hosts, we found that the gaseous satellite numbers, $0-4$, agree across the different studies (Figure \ref{fig:Mg_cut_satellite_count_Rvir_comprehensive}). Between ALFALFA and ELVES (both local volume spiral hosts), we have some overlap in hosts and found overall agreement in the star-forming/gaseous satellite population (Figure \ref{fig:alfa_elves_host_direct_comparison}).


There are several key outcomes of this paper that can be highlighted.
\begin{itemize}
    \item We present a spatial-and-kinematic selection of satellite galaxies, coupled with a statistical quantification of spatial uncertainties using the line-of-sight distances and errors. Of the 5 overlapping hosts between ALFALFA and ELVES, 7/8 ALFALFA satellites within $R_{200}$ are confirmed by ELVES.  This indicates the selection method is working well and can be used for future surveys. 
    \item The Milky Way and M31 are not unusual in terms of only having $<4$ gaseous dwarf galaxy satellites within their virial radii compared to other Local Volume spiral galaxies.  Given the large number of satellites expected under $\Lambda$CDM, this could indicate that satellite stripping mechanisms, such as ram pressure stripping by a halo medium, are efficient in other systems. For our ALFALFA sample with nonzero gaseous satellites, we predict on average $2-7$ quenched satellites within $R_{200}$ at the ELVES limit, but individual host scatter remains non-negligible.
    \item  The properties (spatial and kinematic proximity, HI mass, velocity width) of the candidate gaseous dwarf satellites of other spiral galaxies are overall similar to the properties of the gaseous dwarf galaxies within the Local Group (within the ALFALFA detection limits). For these two populations, we also calculate an HI gas-to-total mass ratio within the gaseous radius and find a consistent value of $M_{\rm HI}$/$M_{\rm dyn} \approx 12 \%$. A contrasting point is that satellites with gas masses as large as the Magellanic Clouds are not detected. 
    \item The number of gaseous satellite galaxies detected here within $R_{200}$ is consistent with the number of star-forming satellite dwarf galaxies detected in the deep optical surveys ELVES and SAGA, as well as in the Auriga cosmological simulations.  We do not see the discrepancy with the Local Group that SAGA sees; however, we have applied an $R_{200}$ cut spatially rather than a uniform 300 kpc cut and excluded potential satellites larger than the LMC.  
    \item There is a moderate increase in gaseous satellite numbers with increasing host halo mass.  This is consistent with the ELVES finding of increasing total satellite number with host luminosity. The ELVES result of a mild quenched fraction trend with host luminosity, however, flattens when the projected area is scaled to the virial radius instead of 300 kpc. 
\end{itemize}

Independent of the trends noted above, surveys of satellites around spiral galaxies have found a small number of `outlier' hosts with a high number of gaseous/star-forming satellites. There is abundant literature characterizing typical versus outlier systems and the underlying galaxy evolution processes. Current directions include expanding the observational sample size, 
assessing the quenched fraction within a physically motivated spatial area (e.g., $R_{200}$ as in Figure \ref{fig:quenched_fraction_spatial_and_luminosity_cut}), checking consistency with cosmological simulations \citep{carlsten_radial_2020,karunakaran_satellites_2021-1,font_can_2021,font_quenching_2022,engler_abundance_2021-1,samuel_extinguishing_2022}, and examining the gaseous halo environments, especially in a galaxy group context \citep{nuza_distribution_2014-1,garrison-kimmel_local_2019,putman_gas_2021,damle_cold_2022-1}. The group environment might have complex effects on gaseous/star-forming satellite abundances: a gaseous group medium could result in satellite ram pressure stripping beyond the individual spiral galaxy's $R_{200}$ \citep{hester_ram_2006,mccarthy_ram_2008,putman_gas_2021}, as well as a more complex tidal or merger history \citep{smercina_relating_2022}.
Observational effects and the chosen analysis methods could also play a significant role, e.g., contamination, group finding process, satellite association definitions \citep{wang_dearth_2018,wang_dearth_2018-1,davies_galaxy_2019}.

The next-generation space telescope observations, such as with the James Webb Space Telescope (JWST), and deep ground-based surveys, such as the Vera Rubin Observatory, have the power to probe the low surface brightness galaxy populations and improve galactic distance measurements. Such future observations will better constrain the satellite-host association and quenched fraction and help obtain key information such as the luminosity, size, and morphology of the dwarfs. From the HI perspective, ongoing and future deep HI surveys such as WALLABY will detect some of the lowest mass dwarf satellites with gas currently missing in our sample. WALLABY's depth and wide sky coverage will also significantly expand the nearby spiral host sample, which enables statistically more representative modeling. Our results for nearby spiral galaxies and their gaseous satellites are important for understanding dwarf galaxy evolution and gas-quenching mechanisms beyond the Local Group.

\section*{Acknowledgements}
We thank Christine Simpson for access to the Auriga simulation data and Marla Geha and Scott Carlsten for additional information on the SAGA and ELVES data.  We also thank the following for useful conversations: Greg Bryan, Jenny Greene, Kathryn Johnston, Stephanie Tonnesen, Tobias Westmeier, and Yong Zheng. We thank the anonymous referee for helpful comments that improved the quality of this paper. This research has made use of the Arecibo Legacy Fast ALFA (ALFALFA) data. We acknowledge the use of the NASA/IPAC Extragalactic Database (NED) and the open-source Python packages Astropy, Numpy, Scipy, Matplotlib, and IPython.

\section*{Data Availability}

The data underlying this article are available in the article and in its online supplementary material.



\bibliographystyle{mnras}
\bibliography{gaseous_dwarfs}

\begin{thebibliography}{}
\makeatletter
\relax
\def\mn@urlcharsother{\let\do\@makeother \do\$\do\&\do\#\do\^\do\_\do\%\do\~}
\def\mn@doi{\begingroup\mn@urlcharsother \@ifnextchar [ {\mn@doi@}
  {\mn@doi@[]}}
\def\mn@doi@[#1]#2{\def\@tempa{#1}\ifx\@tempa\@empty \href
  {http://dx.doi.org/#2} {doi:#2}\else \href {http://dx.doi.org/#2} {#1}\fi
  \endgroup}
\def\mn@eprint#1#2{\mn@eprint@#1:#2::\@nil}
\def\mn@eprint@arXiv#1{\href {http://arxiv.org/abs/#1} {{\tt arXiv:#1}}}
\def\mn@eprint@dblp#1{\href {http://dblp.uni-trier.de/rec/bibtex/#1.xml}
  {dblp:#1}}
\def\mn@eprint@#1:#2:#3:#4\@nil{\def\@tempa {#1}\def\@tempb {#2}\def\@tempc
  {#3}\ifx \@tempc \@empty \let \@tempc \@tempb \let \@tempb \@tempa \fi \ifx
  \@tempb \@empty \def\@tempb {arXiv}\fi \@ifundefined
  {mn@eprint@\@tempb}{\@tempb:\@tempc}{\expandafter \expandafter \csname
  mn@eprint@\@tempb\endcsname \expandafter{\@tempc}}}

\bibitem[\protect\citeauthoryear{Adams \& Oosterloo}{Adams \&
  Oosterloo}{2018}]{adams_deep_2018}
Adams E. A.~K.,  Oosterloo T.~A.,  2018, \aap, 612, A26

\bibitem[\protect\citeauthoryear{Agertz et~al.,}{Agertz
  et~al.}{2013}]{agertz_toward_2013}
Agertz O.,  et~al., 2013, \apj, 770, 25

\bibitem[\protect\citeauthoryear{Ann, Seo  \& Ha}{Ann
  et~al.}{2015}]{2015ApJS..217...27A}
Ann H.~B.,  Seo M.,   Ha D.~K.,  2015, \apjs, 217, 27

\bibitem[\protect\citeauthoryear{Annuar et~al.,}{Annuar
  et~al.}{2020}]{annuar_nustar_2020}
Annuar A.,  et~al., 2020, \mnras, 497, 229

\bibitem[\protect\citeauthoryear{Babul \& Rees}{Babul \&
  Rees}{1992}]{babul_dwarf_1992}
Babul A.,  Rees M.~J.,  1992, \mnras, 255, 346

\bibitem[\protect\citeauthoryear{Bah{\'e} \& McCarthy}{Bah{\'e} \&
  McCarthy}{2015}]{bahe_star_2015}
Bah{\'e} Y.~M.,  McCarthy I.~G.,  2015, \mnras, 447, 969

\bibitem[\protect\citeauthoryear{Baldry et~al.,}{Baldry
  et~al.}{2006}]{baldry_galaxy_2006}
Baldry I.~K.,  et~al., 2006, \mnras, 373, 469

\bibitem[\protect\citeauthoryear{Barnes \& {de Blok}}{Barnes \& {de
  Blok}}{2001}]{barnes_neutral_2001}
Barnes D.~G.,  {de Blok} W. J.~G.,  2001, \aj, 122, 825

\bibitem[\protect\citeauthoryear{Bekki}{Bekki}{2014}]{bekki_galactic_2014}
Bekki K.,  2014, \mnras, 438, 444

\bibitem[\protect\citeauthoryear{{Bland-Hawthorn} \& Gerhard}{{Bland-Hawthorn}
  \& Gerhard}{2016}]{bland-hawthorn_galaxy_2016-1}
{Bland-Hawthorn} J.,  Gerhard O.,  2016, \araa, 54, 529

\bibitem[\protect\citeauthoryear{Bluck et~al.,}{Bluck
  et~al.}{2020}]{bluck_how_2020-1}
Bluck A. F.~L.,  et~al., 2020, \mnras, 499, 230

\bibitem[\protect\citeauthoryear{Boselli \& Gavazzi}{Boselli \&
  Gavazzi}{2014}]{boselli_origin_2014}
Boselli A.,  Gavazzi G.,  2014, \aapr, 22, 74

\bibitem[\protect\citeauthoryear{Boselli et~al.,}{Boselli
  et~al.}{2008}]{boselli_origin_2008}
Boselli A.,  et~al., 2008, \apj, 674, 742

\bibitem[\protect\citeauthoryear{Boselli et~al.,}{Boselli
  et~al.}{2014}]{boselli_cold_2014}
Boselli A.,  et~al., 2014, \aap, 564, A66

\bibitem[\protect\citeauthoryear{Bowen et~al.,}{Bowen
  et~al.}{2016}]{bowen_structure_2016}
Bowen D.~V.,  et~al., 2016, \apj, 826, 50

\bibitem[\protect\citeauthoryear{Brown et~al.,}{Brown
  et~al.}{2017}]{brown_cold_2017}
Brown T.,  et~al., 2017, \mnras, 466, 1275

\bibitem[\protect\citeauthoryear{Cair{\'o}s et~al.,}{Cair{\'o}s
  et~al.}{2001}]{cairos_multiband_2001}
Cair{\'o}s L.~M.,  et~al., 2001, \apjs, 136, 393

\bibitem[\protect\citeauthoryear{Carignan et~al.,}{Carignan
  et~al.}{2013}]{carignan_kat-7_2013}
Carignan C.,  et~al., 2013, \aj, 146, 48

\bibitem[\protect\citeauthoryear{Carlsten et~al.,}{Carlsten
  et~al.}{2020a}]{carlsten_wide-field_2020}
Carlsten S.~G.,  et~al., 2020a, \apj, 891, 144

\bibitem[\protect\citeauthoryear{Carlsten et~al.,}{Carlsten
  et~al.}{2020b}]{carlsten_radial_2020}
Carlsten S.~G.,  et~al., 2020b, \apj, 902, 124

\bibitem[\protect\citeauthoryear{Carlsten et~al.,}{Carlsten
  et~al.}{2021}]{carlsten_luminosity_2021}
Carlsten S.~G.,  et~al., 2021, \apj, 908, 109

\bibitem[\protect\citeauthoryear{Carlsten et~al.,}{Carlsten
  et~al.}{2022}]{carlsten_exploration_2022-1}
Carlsten S.~G.,  et~al., 2022, \apj, 933, 47

\bibitem[\protect\citeauthoryear{Cohen et~al.,}{Cohen
  et~al.}{2018}]{cohen_dragonfly_2018}
Cohen Y.,  et~al., 2018, \apj, 868, 96

\bibitem[\protect\citeauthoryear{Collaboration et~al.,}{Collaboration
  et~al.}{2020}]{collaboration_planck_2020}
Collaboration P.,  et~al., 2020, \aap, 641, A6

\bibitem[\protect\citeauthoryear{Cook et~al.,}{Cook
  et~al.}{2014}]{cook_spitzer_2014}
Cook D.~O.,  et~al., 2014, \mnras, 445, 899

\bibitem[\protect\citeauthoryear{Cora et~al.,}{Cora
  et~al.}{2018}]{cora_semi-analytic_2018}
Cora S.~A.,  et~al., 2018, \mnras, 479, 2

\bibitem[\protect\citeauthoryear{Cortese, Catinella  \& Smith}{Cortese
  et~al.}{2021}]{cortese_dawes_2021}
Cortese L.,  Catinella B.,   Smith R.,  2021, PASA, 38, e035

\bibitem[\protect\citeauthoryear{Croton et~al.,}{Croton
  et~al.}{2006}]{croton_many_2006}
Croton D.~J.,  et~al., 2006, \mnras, 365, 11

\bibitem[\protect\citeauthoryear{Dalla~Vecchia \& Schaye}{Dalla~Vecchia \&
  Schaye}{2008}]{dalla_vecchia_simulating_2008}
Dalla~Vecchia C.,  Schaye J.,  2008, \mnras, 387, 1431

\bibitem[\protect\citeauthoryear{Damle et~al.,}{Damle
  et~al.}{2022}]{damle_cold_2022-1}
Damle M.,  et~al., 2022, \mnras, 512, 3717

\bibitem[\protect\citeauthoryear{Davies et~al.,}{Davies
  et~al.}{2019}]{davies_galaxy_2019}
Davies L. J.~M.,  et~al., 2019, \mnras, 483, 5444

\bibitem[\protect\citeauthoryear{Dicaire et~al.,}{Dicaire
  et~al.}{2008}]{2008MNRAS.385..553D}
Dicaire I.,  et~al., 2008, \mnras, 385, 553

\bibitem[\protect\citeauthoryear{Diemand, Moore  \& Stadel}{Diemand
  et~al.}{2004}]{diemand_velocity_2004}
Diemand J.,  Moore B.,   Stadel J.,  2004, \mnras, 352, 535

\bibitem[\protect\citeauthoryear{Durbala et~al.,}{Durbala
  et~al.}{2020}]{durbala_alfalfa-sdss_2020-1}
Durbala A.,  et~al., 2020, AJ, 160, 271

\bibitem[\protect\citeauthoryear{Emerick et~al.,}{Emerick
  et~al.}{2016}]{emerick_gas_2016}
Emerick A.,  et~al., 2016, \apj, 826, 148

\bibitem[\protect\citeauthoryear{Engler et~al.,}{Engler
  et~al.}{2021}]{engler_abundance_2021-1}
Engler C.,  et~al., 2021, \mnras, 507, 4211

\bibitem[\protect\citeauthoryear{Erwin}{Erwin}{2004}]{erwin_double-barred_2004}
Erwin P.,  2004, \aap, 415, 941

\bibitem[\protect\citeauthoryear{Fabricius et~al.,}{Fabricius
  et~al.}{2012}]{2012ApJ...754...67F}
Fabricius M.~H.,  et~al., 2012, \apj, 754, 67

\bibitem[\protect\citeauthoryear{Font, McCarthy  \& Belokurov}{Font
  et~al.}{2021}]{font_can_2021}
Font A.~S.,  McCarthy I.~G.,   Belokurov V.,  2021, \mnras, 505, 783

\bibitem[\protect\citeauthoryear{Font et~al.,}{Font
  et~al.}{2022}]{font_quenching_2022}
Font A.~S.,  et~al., 2022, \mnras, 511, 1544

\bibitem[\protect\citeauthoryear{Fouque et~al.,}{Fouque
  et~al.}{1992}]{fouque_groups_1992}
Fouque P.,  et~al., 1992, \aaps, 93, 211

\bibitem[\protect\citeauthoryear{Garcia}{Garcia}{1993}]{garcia_general_1993}
Garcia A.~M.,  1993, \aaps, 100, 47

\bibitem[\protect\citeauthoryear{{Garrison-Kimmel} et~al.,}{{Garrison-Kimmel}
  et~al.}{2019}]{garrison-kimmel_local_2019}
{Garrison-Kimmel} S.,  et~al., 2019, \mnras, 487, 1380

\bibitem[\protect\citeauthoryear{Geha et~al.,}{Geha
  et~al.}{2012}]{geha_stellar_2012}
Geha M.,  et~al., 2012, \apj, 757, 85

\bibitem[\protect\citeauthoryear{Geha et~al.,}{Geha
  et~al.}{2017}]{geha_saga_2017}
Geha M.,  et~al., 2017, \apj, 847, 4

\bibitem[\protect\citeauthoryear{Giovanelli et~al.,}{Giovanelli
  et~al.}{2005}]{giovanelli_arecibo_2005}
Giovanelli R.,  et~al., 2005, \aj, 130, 2613

\bibitem[\protect\citeauthoryear{Gnedin}{Gnedin}{2000}]{gnedin_cosmological_2000}
Gnedin N.~Y.,  2000, \apj, 535, 530

\bibitem[\protect\citeauthoryear{Grand et~al.,}{Grand
  et~al.}{2017}]{grand_auriga_2017-1}
Grand R. J.~J.,  et~al., 2017, \mnras, 467, 179

\bibitem[\protect\citeauthoryear{Grcevich \& Putman}{Grcevich \&
  Putman}{2009}]{grcevich_h_2009}
Grcevich J.,  Putman M.~E.,  2009, \apj, 696, 385

\bibitem[\protect\citeauthoryear{Gregory \& Thompson}{Gregory \&
  Thompson}{1977}]{1977ApJ...213..345G}
Gregory S.~A.,  Thompson L.~A.,  1977, \apj, 213, 345

\bibitem[\protect\citeauthoryear{Grossi et~al.,}{Grossi
  et~al.}{2009}]{grossi_hi_2009}
Grossi M.,  et~al., 2009, \aap, 498, 407

\bibitem[\protect\citeauthoryear{Gunn \& Gott}{Gunn \&
  Gott}{1972}]{gunn_infall_1972-1}
Gunn J.~E.,  Gott III J.~R.,  1972, \apj, 176, 1

\bibitem[\protect\citeauthoryear{Guo et~al.,}{Guo
  et~al.}{2012}]{guo_satellite_2012}
Guo Q.,  et~al., 2012, \mnras, 427, 428

\bibitem[\protect\citeauthoryear{Haynes}{Haynes}{1979}]{haynes_neutral_1979}
Haynes M.~P.,  1979, \aj, 84, 1830

\bibitem[\protect\citeauthoryear{Haynes et~al.,}{Haynes
  et~al.}{2011}]{haynes_arecibo_2011}
Haynes M.~P.,  et~al., 2011, \aj, 142, 170

\bibitem[\protect\citeauthoryear{Haynes et~al.,}{Haynes
  et~al.}{2018}]{haynes_arecibo_2018}
Haynes M.~P.,  et~al., 2018, \apj, 861, 49

\bibitem[\protect\citeauthoryear{Hester}{Hester}{2006}]{hester_ram_2006}
Hester J.~A.,  2006, \apj, 647, 910

\bibitem[\protect\citeauthoryear{Hester et~al.,}{Hester
  et~al.}{2010}]{hester_ic_2010}
Hester J.~A.,  et~al., 2010, \apj, 716, L14

\bibitem[\protect\citeauthoryear{Ho, Filippenko  \& Sargent}{Ho
  et~al.}{1997}]{ho_search_1997}
Ho L.~C.,  Filippenko A.~V.,   Sargent W. L.~W.,  1997, \apjs, 112, 315

\bibitem[\protect\citeauthoryear{Hopkins et~al.,}{Hopkins
  et~al.}{2018}]{hopkins_fire-2_2018}
Hopkins P.~F.,  et~al., 2018, \mnras, 480, 800

\bibitem[\protect\citeauthoryear{Hunter et~al.,}{Hunter
  et~al.}{1989}]{hunter_star_1989}
Hunter D.~A.,  et~al., 1989, \apj, 341, 697

\bibitem[\protect\citeauthoryear{Irwin et~al.,}{Irwin
  et~al.}{2009}]{irwin_cdm_2009}
Irwin J.~A.,  et~al., 2009, \apj, 692, 1447

\bibitem[\protect\citeauthoryear{Jones et~al.,}{Jones
  et~al.}{2018}]{jones_alfalfa_2018}
Jones M.~G.,  et~al., 2018, \mnras, 477, 2

\bibitem[\protect\citeauthoryear{Karachentsev, Makarov  \&
  Kaisina}{Karachentsev et~al.}{2013}]{karachentsev_updated_2013}
Karachentsev I.~D.,  Makarov D.~I.,   Kaisina E.~I.,  2013, \aj, 145, 101

\bibitem[\protect\citeauthoryear{Karachentsev et~al.,}{Karachentsev
  et~al.}{2014}]{karachentsev_infall_2014}
Karachentsev I.~D.,  et~al., 2014, \apj, 782, 4

\bibitem[\protect\citeauthoryear{Karachentsev et~al.,}{Karachentsev
  et~al.}{2015}]{karachentsev_peculiar_2015}
Karachentsev I.~D.,  et~al., 2015, \apj, 805, 144

\bibitem[\protect\citeauthoryear{Karachentsev, Kaisina  \&
  Makarov}{Karachentsev et~al.}{2018}]{karachentsev_morphological_2018}
Karachentsev I.~D.,  Kaisina E.~I.,   Makarov D.~I.,  2018, \mnras, 479, 4136

\bibitem[\protect\citeauthoryear{Karunakaran et~al.,}{Karunakaran
  et~al.}{2021}]{karunakaran_satellites_2021-1}
Karunakaran A.,  et~al., 2021, \apj, 916, L19

\bibitem[\protect\citeauthoryear{Kauffmann et~al.,}{Kauffmann
  et~al.}{2004}]{kauffmann_environmental_2004}
Kauffmann G.,  et~al., 2004, \mnras, 353, 713

\bibitem[\protect\citeauthoryear{Kim et~al.,}{Kim
  et~al.}{2020}]{2020ApJ...905..104K}
Kim Y.~J.,  et~al., 2020, \apj, 905, 104

\bibitem[\protect\citeauthoryear{Kondapally et~al.,}{Kondapally
  et~al.}{2018}]{kondapally_faint_2018}
Kondapally R.,  et~al., 2018, \mnras, 481, 1759

\bibitem[\protect\citeauthoryear{Koribalski et~al.,}{Koribalski
  et~al.}{2020}]{koribalski_wallaby_2020}
Koribalski B.~S.,  et~al., 2020, \apss, 365, 118

\bibitem[\protect\citeauthoryear{Kourkchi \& Tully}{Kourkchi \&
  Tully}{2017}]{kourkchi_galaxy_2017}
Kourkchi E.,  Tully R.~B.,  2017, \apj, 843, 16

\bibitem[\protect\citeauthoryear{Kourkchi et~al.,}{Kourkchi
  et~al.}{2020}]{kourkchi_cosmicflows-4_2020}
Kourkchi E.,  et~al., 2020, \apj, 902, 145

\bibitem[\protect\citeauthoryear{Liu et~al.,}{Liu et~al.}{2011}]{liu_how_2011}
Liu L.,  et~al., 2011, \apj, 733, 62

\bibitem[\protect\citeauthoryear{Mac~Low \& Klessen}{Mac~Low \&
  Klessen}{2004}]{mac_low_control_2004}
Mac~Low M.-M.,  Klessen R.~S.,  2004, Rev. Mod. Phys., 76, 125

\bibitem[\protect\citeauthoryear{Makarov \& Karachentsev}{Makarov \&
  Karachentsev}{2011}]{2011MNRAS.412.2498M}
Makarov D.,  Karachentsev I.,  2011, \mnras, 412, 2498

\bibitem[\protect\citeauthoryear{Mao et~al.,}{Mao et~al.}{2021}]{mao_saga_2021}
Mao Y.-Y.,  et~al., 2021, \apj, 907, 85

\bibitem[\protect\citeauthoryear{Mateo}{Mateo}{1998}]{mateo_dwarf_1998}
Mateo M.~L.,  1998, \araa, 36, 435

\bibitem[\protect\citeauthoryear{Mateo, Olszewski  \& Walker}{Mateo
  et~al.}{2008}]{mateo_velocity_2008}
Mateo M.,  Olszewski E.~W.,   Walker M.~G.,  2008, \apj, 675, 201

\bibitem[\protect\citeauthoryear{Mayer et~al.,}{Mayer
  et~al.}{2001}]{mayer_tidal_2001}
Mayer L.,  et~al., 2001, \apj, 547, L123

\bibitem[\protect\citeauthoryear{Mayer et~al.,}{Mayer
  et~al.}{2006}]{mayer_simultaneous_2006}
Mayer L.,  et~al., 2006, \mnras, 369, 1021

\bibitem[\protect\citeauthoryear{McCarthy et~al.,}{McCarthy
  et~al.}{2008}]{mccarthy_ram_2008}
McCarthy I.~G.,  et~al., 2008, \mnras, 383, 593

\bibitem[\protect\citeauthoryear{McConnachie}{McConnachie}{2012}]{mcconnachie_observed_2012}
McConnachie A.~W.,  2012, \aj, 144, 4

\bibitem[\protect\citeauthoryear{McGaugh \& van Dokkum}{McGaugh \& van
  Dokkum}{2021}]{mcgaugh_dark_2021}
McGaugh S.~S.,  van Dokkum P.,  2021, Res. Notes AAS, 5, 23

\bibitem[\protect\citeauthoryear{McKee \& Ostriker}{McKee \&
  Ostriker}{2007}]{mckee_theory_2007}
McKee C.~F.,  Ostriker E.~C.,  2007, \araa, 45, 565

\bibitem[\protect\citeauthoryear{McQuinn et~al.,}{McQuinn
  et~al.}{2017}]{2017AJ....154...51M}
McQuinn K. B.~W.,  et~al., 2017, \aj, 154, 51

\bibitem[\protect\citeauthoryear{Meurer et~al.,}{Meurer
  et~al.}{2006}]{meurer_survey_2006}
Meurer G.~R.,  et~al., 2006, \apjs, 165, 307

\bibitem[\protect\citeauthoryear{Moster et~al.,}{Moster
  et~al.}{2010}]{moster_constraints_2010}
Moster B.~P.,  et~al., 2010, \apj, 710, 903

\bibitem[\protect\citeauthoryear{M{\"u}ller, Jerjen  \& Binggeli}{M{\"u}ller
  et~al.}{2018}]{muller_leo-i_2018}
M{\"u}ller O.,  Jerjen H.,   Binggeli B.,  2018, \aap, 615, A105

\bibitem[\protect\citeauthoryear{Nelson et~al.,}{Nelson
  et~al.}{2019}]{nelson_first_2019}
Nelson D.,  et~al., 2019, \mnras, 490, 3234

\bibitem[\protect\citeauthoryear{Noordermeer et~al.,}{Noordermeer
  et~al.}{2005}]{noordermeer_westerbork_2005}
Noordermeer E.,  et~al., 2005, \aap, 442, 137

\bibitem[\protect\citeauthoryear{Nowak et~al.,}{Nowak
  et~al.}{2010}]{2010MNRAS.403..646N}
Nowak N.,  et~al., 2010, \mnras, 403, 646

\bibitem[\protect\citeauthoryear{Nuza et~al.,}{Nuza
  et~al.}{2014}]{nuza_distribution_2014-1}
Nuza S.~E.,  et~al., 2014, \mnras, 441, 2593

\bibitem[\protect\citeauthoryear{Paturel et~al.,}{Paturel
  et~al.}{2003}]{paturel_hyperleda_2003}
Paturel G.,  et~al., 2003, \aap, 412, 45

\bibitem[\protect\citeauthoryear{Pearson et~al.,}{Pearson
  et~al.}{2016}]{pearson_local_2016}
Pearson S.,  et~al., 2016, \mnras, 459, 1827

\bibitem[\protect\citeauthoryear{Peng et~al.,}{Peng
  et~al.}{2010}]{peng_mass_2010}
Peng Y.-j.,  et~al., 2010, \apj, 721, 193

\bibitem[\protect\citeauthoryear{Peng et~al.,}{Peng
  et~al.}{2012}]{peng_mass_2012}
Peng Y.-j.,  et~al., 2012, \apj, 757, 4

\bibitem[\protect\citeauthoryear{Phillips et~al.,}{Phillips
  et~al.}{2014}]{phillips_dichotomy_2014}
Phillips J.~I.,  et~al., 2014, \mnras, 437, 1930

\bibitem[\protect\citeauthoryear{Pillepich et~al.,}{Pillepich
  et~al.}{2019}]{pillepich_first_2019}
Pillepich A.,  et~al., 2019, \mnras, 490, 3196

\bibitem[\protect\citeauthoryear{Pisano et~al.,}{Pisano
  et~al.}{2007}]{pisano_h_2007}
Pisano D.~J.,  et~al., 2007, \apj, 662, 959

\bibitem[\protect\citeauthoryear{Putman et~al.,}{Putman
  et~al.}{2021}]{putman_gas_2021}
Putman M.~E.,  et~al., 2021, \apj, 913, 53

\bibitem[\protect\citeauthoryear{Qu, Bregman  \& {Hodges-Kluck}}{Qu
  et~al.}{2019}]{qu_hstcos_2019}
Qu Z.,  Bregman J.~N.,   {Hodges-Kluck} E.~J.,  2019, \apj, 876, 101

\bibitem[\protect\citeauthoryear{Robotham et~al.,}{Robotham
  et~al.}{2012}]{robotham_galaxy_2012}
Robotham A. S.~G.,  et~al., 2012, \mnras, 424, 1448

\bibitem[\protect\citeauthoryear{Rodriguez~Wimberly et~al.,}{Rodriguez~Wimberly
  et~al.}{2019}]{rodriguez_wimberly_suppression_2019}
Rodriguez~Wimberly M.~K.,  et~al., 2019, \mnras, 483, 4031

\bibitem[\protect\citeauthoryear{Roediger \& Hensler}{Roediger \&
  Hensler}{2005}]{roediger_ram_2005}
Roediger E.,  Hensler G.,  2005, \aap, 433, 875

\bibitem[\protect\citeauthoryear{Sabbi et~al.,}{Sabbi
  et~al.}{2018}]{sabbi_resolved_2018}
Sabbi E.,  et~al., 2018, \apjs, 235, 23

\bibitem[\protect\citeauthoryear{Samuel et~al.,}{Samuel
  et~al.}{2022}]{samuel_extinguishing_2022}
Samuel J.,  et~al., 2022, \mnras, 514, 5276

\bibitem[\protect\citeauthoryear{Schneider et~al.,}{Schneider
  et~al.}{1983}]{1983ApJ...273L...1S}
Schneider S.~E.,  et~al., 1983, \apj, 273, L1

\bibitem[\protect\citeauthoryear{Schombert \& McGaugh}{Schombert \&
  McGaugh}{2021}]{schombert_anomalous_2021}
Schombert J.,  McGaugh S.,  2021, \aj, 161, 91

\bibitem[\protect\citeauthoryear{Shaw et~al.,}{Shaw
  et~al.}{1995}]{1995MNRAS.274..369S}
Shaw M.,  et~al., 1995, \mnras, 274, 369

\bibitem[\protect\citeauthoryear{Shaya et~al.,}{Shaya
  et~al.}{2017}]{shaya_action_2017}
Shaya E.~J.,  et~al., 2017, \apj, 850, 207

\bibitem[\protect\citeauthoryear{Simon}{Simon}{2019}]{simon_faintest_2019-1}
Simon J.~D.,  2019, \araa, 57, 375

\bibitem[\protect\citeauthoryear{Simpson et~al.,}{Simpson
  et~al.}{2018}]{simpson_quenching_2018}
Simpson C.~M.,  et~al., 2018, \mnras, 478, 548

\bibitem[\protect\citeauthoryear{Smercina et~al.,}{Smercina
  et~al.}{2018}]{smercina_lonely_2018}
Smercina A.,  et~al., 2018, ApJ, 863, 152

\bibitem[\protect\citeauthoryear{Smercina et~al.,}{Smercina
  et~al.}{2020}]{smercina_saga_2020}
Smercina A.,  et~al., 2020, \apj, 905, 60

\bibitem[\protect\citeauthoryear{Smercina et~al.,}{Smercina
  et~al.}{2022}]{smercina_relating_2022}
Smercina A.,  et~al., 2022, \apj, 930, 69

\bibitem[\protect\citeauthoryear{Somerville}{Somerville}{2002}]{somerville_can_2002}
Somerville R.~S.,  2002, \apj, 572, L23

\bibitem[\protect\citeauthoryear{Spekkens et~al.,}{Spekkens
  et~al.}{2014}]{spekkens_dearth_2014}
Spekkens K.,  et~al., 2014, \apj, 795, L5

\bibitem[\protect\citeauthoryear{Spencer, Loebman  \& Yoachim}{Spencer
  et~al.}{2014}]{spencer_survey_2014}
Spencer M.,  Loebman S.,   Yoachim P.,  2014, \apj, 788, 146

\bibitem[\protect\citeauthoryear{Springel et~al.,}{Springel
  et~al.}{2008}]{springel_aquarius_2008}
Springel V.,  et~al., 2008, \mnras, 391, 1685

\bibitem[\protect\citeauthoryear{{Staveley-Smith} et~al.,}{{Staveley-Smith}
  et~al.}{2003}]{staveley-smith_new_2003}
{Staveley-Smith} L.,  et~al., 2003, \mnras, 339, 87

\bibitem[\protect\citeauthoryear{Steyrleithner, Hensler  \&
  Boselli}{Steyrleithner et~al.}{2020}]{steyrleithner_effect_2020}
Steyrleithner P.,  Hensler G.,   Boselli A.,  2020, \mnras, 494, 1114

\bibitem[\protect\citeauthoryear{Stierwalt et~al.,}{Stierwalt
  et~al.}{2009}]{stierwalt_arecibo_2009}
Stierwalt S.,  et~al., 2009, AJ, 138, 338

\bibitem[\protect\citeauthoryear{Swaters et~al.,}{Swaters
  et~al.}{2002}]{swaters_westerbork_2002}
Swaters R.~A.,  et~al., 2002, \aap, 390, 829

\bibitem[\protect\citeauthoryear{Tanaka et~al.,}{Tanaka
  et~al.}{2018}]{tanaka_missing_2018}
Tanaka M.,  et~al., 2018, ApJ, 865, 125

\bibitem[\protect\citeauthoryear{Tchernyshyov et~al.,}{Tchernyshyov
  et~al.}{2022}]{tchernyshyov_cgm2_2022}
Tchernyshyov K.,  et~al., 2022, \apj, 927, 147

\bibitem[\protect\citeauthoryear{Tollerud et~al.,}{Tollerud
  et~al.}{2011}]{tollerud_small-scale_2011}
Tollerud E.~J.,  et~al., 2011, \apj, 738, 102

\bibitem[\protect\citeauthoryear{Tonnesen \& Bryan}{Tonnesen \&
  Bryan}{2012}]{tonnesen_star_2012}
Tonnesen S.,  Bryan G.~L.,  2012, \mnras, 422, 1609

\bibitem[\protect\citeauthoryear{Trentham \& Tully}{Trentham \&
  Tully}{2009}]{trentham_dwarf_2009}
Trentham N.,  Tully R.~B.,  2009, \mnras, 398, 722

\bibitem[\protect\citeauthoryear{Truong et~al.,}{Truong
  et~al.}{2017}]{truong_high-resolution_2017}
Truong P.~N.,  et~al., 2017, \apj, 843, 37

\bibitem[\protect\citeauthoryear{Tully}{Tully}{1980}]{tully_nearby_1980}
Tully R.~B.,  1980, \apj, 237, 390

\bibitem[\protect\citeauthoryear{Tully}{Tully}{1988}]{1988ngc..book.....T}
Tully R.~B.,  1988, Nearby Galaxies Catalog.
Cambridge: University Press

\bibitem[\protect\citeauthoryear{Tully et~al.,}{Tully
  et~al.}{2009}]{tully_extragalactic_2009}
Tully R.~B.,  et~al., 2009, AJ, 138, 323

\bibitem[\protect\citeauthoryear{Turner \& Gott}{Turner \&
  Gott}{1976}]{1976ApJS...32..409T}
Turner E.~L.,  Gott III J.~R.,  1976, \apjs, 32, 409

\bibitem[\protect\citeauthoryear{Van~Sistine et~al.,}{Van~Sistine
  et~al.}{2016}]{van_sistine_alfalfa_2016}
Van~Sistine A.,  et~al., 2016, \apj, 824, 25

\bibitem[\protect\citeauthoryear{Vargas et~al.,}{Vargas
  et~al.}{2017}]{vargas_halogas_2017}
Vargas C.~J.,  et~al., 2017, ApJ, 839, 118

\bibitem[\protect\citeauthoryear{Wakker et~al.,}{Wakker
  et~al.}{2003}]{wakker_far_2003}
Wakker B.~P.,  et~al., 2003, \apjs, 146, 1

\bibitem[\protect\citeauthoryear{Walker et~al.,}{Walker
  et~al.}{2009}]{walker_universal_2009}
Walker M.~G.,  et~al., 2009, \apj, 704, 1274

\bibitem[\protect\citeauthoryear{Wang et~al.,}{Wang
  et~al.}{2018a}]{wang_dearth_2018}
Wang E.,  et~al., 2018a, \apj, 860, 102

\bibitem[\protect\citeauthoryear{Wang et~al.,}{Wang
  et~al.}{2018b}]{wang_dearth_2018-1}
Wang E.,  et~al., 2018b, \apj, 864, 51

\bibitem[\protect\citeauthoryear{Weisz et~al.,}{Weisz
  et~al.}{2014}]{weisz_star_2014}
Weisz D.~R.,  et~al., 2014, \apj, 789, 148

\bibitem[\protect\citeauthoryear{Westmeier et~al.,}{Westmeier
  et~al.}{2022}]{westmeier_wallaby_2022}
Westmeier T.,  et~al., 2022, PASA, 39, e058

\bibitem[\protect\citeauthoryear{Wetzel, Tinker  \& Conroy}{Wetzel
  et~al.}{2012}]{wetzel_galaxy_2012}
Wetzel A.~R.,  Tinker J.~L.,   Conroy C.,  2012, \mnras, 424, 232

\bibitem[\protect\citeauthoryear{Wetzel, Tollerud  \& Weisz}{Wetzel
  et~al.}{2015}]{2015ApJ...808L..27W}
Wetzel A.~R.,  Tollerud E.~J.,   Weisz D.~R.,  2015, \apj, 808, L27

\bibitem[\protect\citeauthoryear{Young \& Lo}{Young \&
  Lo}{1997a}]{young_neutral_1997}
Young L.~M.,  Lo K.~Y.,  1997a, ApJ, 476, 127

\bibitem[\protect\citeauthoryear{Young \& Lo}{Young \&
  Lo}{1997b}]{1997ApJ...490..710Y}
Young L.~M.,  Lo K.~Y.,  1997b, \apj, 490, 710

\bibitem[\protect\citeauthoryear{{de los Reyes} \& Kennicutt}{{de los Reyes} \&
  Kennicutt}{2019}]{de_los_reyes_revisiting_2019}
{de los Reyes} M. A.~C.,  Kennicutt Jr. R.~C.,  2019, \apj, 872, 16

\makeatother
\end{thebibliography}




\appendix

\section{Likelihood of Three-dimensional Proximity}\label{appendix:Fd3d}
\begin{figure}
    \centering
    \includegraphics[width=1.0\linewidth]{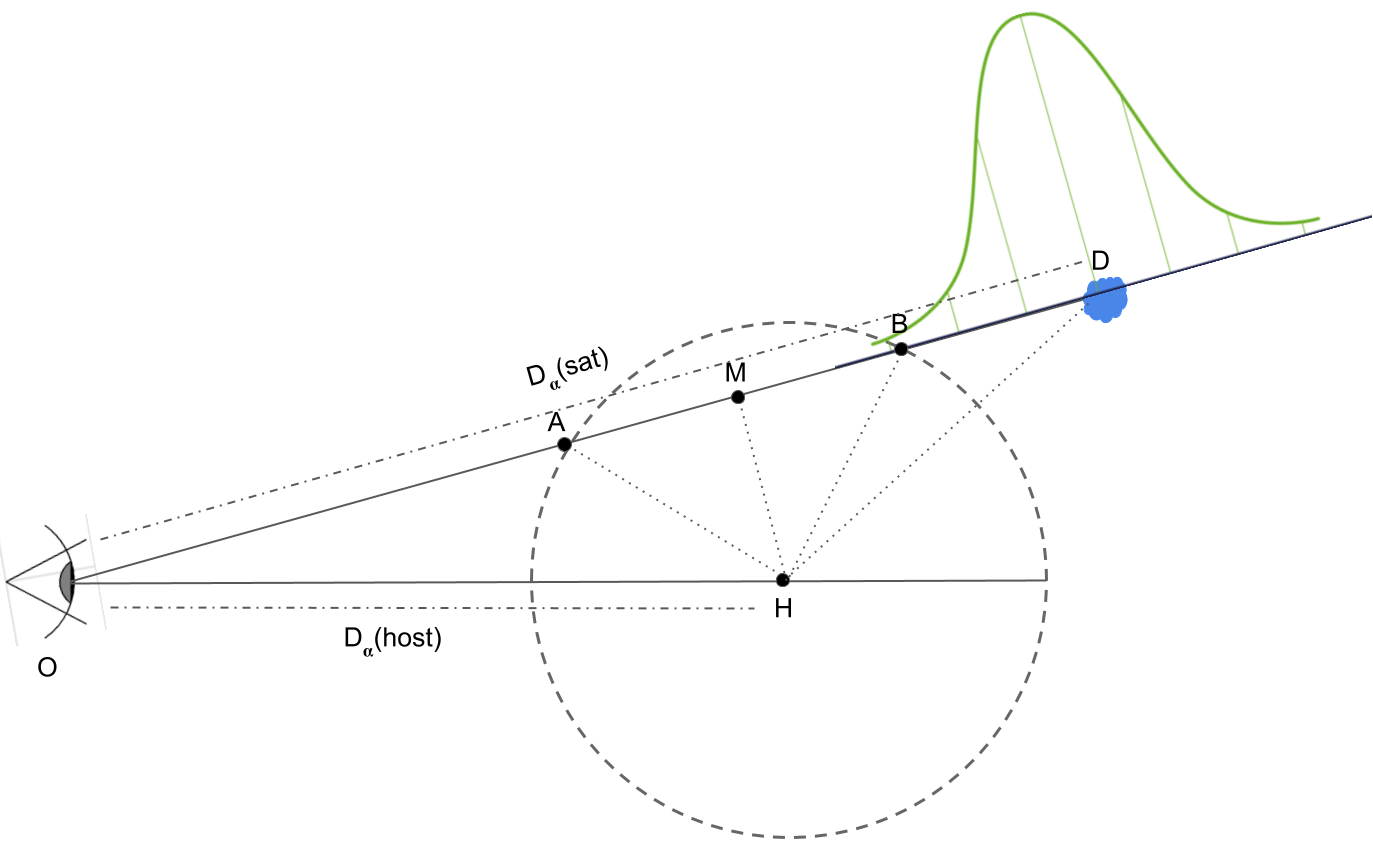}
    \caption{A cartoon illustrating the satellite-host 3D proximity factor ($F_{\rm d_{\rm 3D}}$; equation \ref{eqn:Fd3d_detailed_with_cartoon}). We annotate some useful locations, e.g., the observer `O', the host galaxy `H', and the dwarf satellite candidate `D'. The dashed circle indicates a sphere that encompasses all hosts' $2R_{200}$ halos, here chosen to be $1$ Mpc in radius ($\overline{HA}=\overline{HB}=1$ Mpc). The ALFALFA host ($\overline{OH}$) and satellite ($\overline{OD}$) distances are shown as $D_{\alpha}$(host) and $D_{\alpha}$(sat), respectively. We assume a Gaussian shape for the satellite's line-of-sight distance distribution \citep{jones_alfalfa_2018}, see Appendix \ref{appendix:Fd3d}.}
    \label{fig:psat_cartoon}
\end{figure}

To calculate the probability that a satellite resides within a certain 3D distance to the host, we assume all satellites have a normal distribution\footnote{The ALFALFA catalog distance uncertainties are symmetric. In the case of an asymmetric distance error, the skewed normal distribution should be adopted instead.} of line-of-sight distances, $N(D_{\alpha},\sigma_{D})$, centered at the ALFALFA catalog distance $D_{\alpha}$ with a standard deviation of $\sigma_{D}$ \citep{jones_alfalfa_2018, haynes_arecibo_2018}. 
As illustrated in Figure \ref{fig:psat_cartoon}, the $F_{\rm d_{\rm 3D}}$ factor is the integrated probability that a normal distribution profile (centered at the dwarf `D') falls within a certain radius of the host ($\overline{AB}$): 
\begin{equation}\label{eqn:Fd3d_detailed_with_cartoon}
    F_{\rm d_{\rm 3D}} \coloneqq  P(\rm d_{3D} \leq 1 Mpc) = \int_{-\overline{DA}}^{-\overline{DB}} N(0,\sigma_{D})~dx,
\end{equation}
where $\overline{DA}$ and $\overline{DB}$ can be obtained geometrically, as the angle $\angle DOH$ is the host-satellite angular distance (so that $\overline{MH}=D_{\alpha, \rm host} \cdot \sin(\angle DOH)$), and $\overline{HA}=\overline{HB}=1$ Mpc is a constant radius threshold by choice (see below).

This 3D proximity factor incorporates the line-of-sight distances and uncertainties from the ALFALFA catalog, which supplements the projected distance and velocity-based satellite definitions (\S \ref{subsec:sat_search}). This factor is designed to compare the satellite candidates in terms of the relative spatial proximity to the host (Figure \ref{fig:dvel_dproj_scatter}), but the numerical values ($F_{\rm d_{\rm 3D}} = 25 \pm 14\%$; see Table \ref{table:2R200_satellites}) are relatively low, even for the 7/8 ALFALFA satellites confirmed by deep optical survey (\S \ref{subsec:sf_sats_elves_saga}, Figure \ref{fig:alfa_elves_host_direct_comparison}), suggesting that current distance data are insufficient in determining a satellite candidate's 3D separation to the host within a chosen 1 Mpc spherical volume.

This factor can be used in future targeted optical work, where higher distance accuracy gives higher confidence estimations. This may distinguish true satellite versus field contamination in a more deterministic way. For example, deep optical observations (e.g., \citealt{cohen_dragonfly_2018,2020ApJ...905..104K,carlsten_luminosity_2021}) can reduce a dwarf galaxy's $\sigma_{D}$ in equation \ref{eqn:Fd3d_detailed_with_cartoon} from $\sigma_{D} \approx 2$ to sometimes $\lesssim 1$ Mpc; depending on the distance, this could help confirm or reject the galaxy as a satellite. This is the case for the two NGC 4826 satellite candidates within $R_{200}$, where ELVES confirms our $V_{300}$ object (IC 3840=AGC 229386) and rejects the $V_{\rm esc}$ object (AGC 742601). Using the updated TRGB distances with high accuracy (\citealt{2020ApJ...905..104K}; adopted by ELVES) for the two, $D_{\rm TRGB}=5.86 \pm 0.22$ (IC 3840) and $6.42 \pm 0.32$ Mpc (AGC 742601), the $F_{\rm d_{\rm 3D}}$ values update from $19.2 \%$ to $97.5 \%$, and from $23.6 \%$ to $33.7 \%$ (both scaled to ELVES host distance), respectively. 
At this level of distance accuracy ($\sigma_{D} \approx 0.3$ Mpc), $F_{\rm d_{\rm 3D}}$ values can help determine satellite association to high confidence ($97.5 \%$ for IC 3840).

\section{Results for Individual Host Galaxies}\label{sec:results_individual_hosts}
In this section, we describe the characteristics of each spiral host galaxy and discuss the properties of the satellite candidates within $2R_{200}$ (Table \ref{table:2R200_satellites}) when present. We note that the information for each galaxy varies significantly and is likely incomplete. We refer to the galaxies in the context of their stellar and total (halo) masses alternatively below, where the conversion based on abundance matching is available in Table \ref{table:analog_properties}. When there are overlapping dark matter halos (Figure \ref{fig:virgo_field_sky_distribution}), we put the galaxies together and note the group they are part of.

\subsection{NGC 925}\label{subsec:ngc925}
NGC 925 is a late-type SBcd galaxy with a bright optical and H-$\alpha$ bar. Its ALFALFA distance, $9.1$ Mpc, agrees with the NED average result of $9.16$ Mpc, which our stellar mass reference (Table \ref{table:analog_properties}) adopts. Its projected $R_{200}$ area of $\sim 1{\degr}$ is fully covered in the ALFALFA sky but is relatively close to the $\delta < 36{\degr}$ survey edge. NGC 925 has one of the largest rms noises among our host sample $\sigma_{S21} = 4.01$ mJy, lowering its local HI detectability from $M_{\rm HI} \approx 10^{6.8}$ $M_{\odot}$ (Figure \ref{fig:survey_sensitivity_with_count}) to about $10^{7.0}$ $M_{\odot}$. We find UGC 1924, a relatively massive Scd galaxy, within its $2R_{200}$, and no satellites within its $R_{200}$. NGC 925 is the third brightest galaxy in the gravitationally-bound NGC 1023 group, but the nearest large spiral is a few Mpc away \citep{tully_nearby_1980,pisano_h_2007,trentham_dwarf_2009}. Most of the NGC 1023 group region is outside the ALFALFA sky coverage.

\subsection{NGC 2903}
NGC 2903 is an isolated SAB(rs)bc galaxy. With a derived halo mass of $M_{\rm Halo} \sim 1.6 \times 10^{12}$ $M_{\odot}$, it is one of the closest MW/M31 analogs (by mass) in our sample. All satellite selection methods defined in \S\ref{subsec:sat_search} converge to one gaseous satellite candidate, AGC 191706, with relatively high $F_{\rm d_{\rm 3D}}$ confidence. AGC 191706 is confirmed as a satellite by \citet{irwin_cdm_2009} (deep targeted HI) and \citet{carlsten_exploration_2022-1} (ELVES; deep optical), although both works find additional low-mass HI sources that are likely below the ALFALFA detection limit. In particular, ELVES finds three more gaseous/star-forming satellites and two more quenched satellites within $R_{200}$, which gives a $33 \%$ -- relatively low quenched fraction compared to the MW/M31 (\S \ref{sec:elves_saga_discussion}).

\subsection{NGC 3274}
NGC 3274 is a relatively faint SAB galaxy and the second least massive spiral galaxy in our host sample. We adopt the updated TRGB distance of $D=10$ Mpc \citep{sabbi_resolved_2018} instead of the ALFALFA $D_{\alpha}=6.5$ Mpc value. AGC 731457 is an $R_{200}$ satellite candidate with a stellar mass $M_{*} \approx 9.67 \times 10^{6}$ $M_{\odot}$ from the ALFALFA-SDSS crossmatch \citep{durbala_alfalfa-sdss_2020-1}.  The brightest star distance $D_{BS}=10.52 \pm 0.34$ Mpc for this dwarf galaxy \citep{karachentsev_peculiar_2015} roughly agrees with the ALFALFA value $D_{\alpha}=11.1$ Mpc. The combined projected and line-of-sight proximity gives a relatively high confidence that AGC 731457 is a satellite of NGC 3274 (Table \ref{table:2R200_satellites}). Within $2R_{200}$, we find the less massive dwarf AGC 749315 ($M_{*} \approx 5.18 \times 10^{5}$ $M_{\odot}$; \citealt{durbala_alfalfa-sdss_2020-1}).

\subsection{NGC 3344}
NGC 3344 is a relatively isolated SAB(r)bc galaxy. Similar to NGC 3274, we adopt the updated TRGB distance $D=8.3$ Mpc \citep{sabbi_resolved_2018} in replacement of $D_{\alpha} = 9.8$ Mpc. All of our satellite search methods consistently found zero candidates for NGC 3344, indicating that its vicinity is lacking gas-containing satellites at ALFALFA detectability, which at a higher than average rms beam noise, $\sigma_{S21} = 3.71$ mJy, decreases from approximately $10^{6.7}$ $M_{\odot}$ to $10^{6.9}$ $M_{\odot}$ locally (similar to \S \ref{subsec:ngc925}). This result is confirmed by ELVES, which also finds zero gaseous satellites within a $300$ kpc coverage (comparable to $2R_{200}$; see \S \ref{sec:elves_saga_discussion}).

\subsection{ NGC 3351 and 3368 (Leo I Group)} \label{subsec:LeoI}
NGC 3351 (M95) and NGC 3368 (M96) both belong to the Leo I (M96) Group and have a significant overlap in their projected $R_{200}$ halos (Figure \ref{fig:virgo_field_sky_distribution}). This pair is separated by $611$ kpc in 3D (Table \ref{table:analog_properties}), which is comparable to the MW and M31 separation of $\sim 780$ kpc. We have excluded the HI-only Leo Ring features (S09, \citealt{1983ApJ...273L...1S}) in this field from our dwarf satellite sample. We find that collectively, the Leo I pair hosts $3-4$ satellite candidates in $R_{200}$ and $9$ in $2R_{200}$, and we assign the satellites in the overlapping halo region to the host with a smaller dimensionless separation $\Theta$ (equation \ref{eqn:dimensionless-separation}). The Dragonfly Telephoto Array survey \citep{cohen_dragonfly_2018} mapped 11 LSB objects in the NGC3384/M96 $\sim 3{\degr} \times 3{\degr}$ field, among which an $R_{200}$ satellite of M95: AGC 201970 (LeG 18, see their M96-DF6), is the only overlap with our sample and is Dragonfly-identified as a somewhat irregular ultra diffuse dwarf.

NGC 3351 (M95) is an SB(r)b galaxy with a star-forming H-$\alpha$ ring \citep{2008MNRAS.385..553D}. AGC 200532 (LeG 04) is an $R_{200}$ object distinctively more associated with M95 than M96. AGC 201970 (LeG 18) is M95's $2R_{200}$ object with TRGB and SBF distances of $D_{\rm TRGB} = 10.2 \pm 0.3$ Mpc and $D_{\rm SBF} = 11.3 \pm 3.3$ Mpc \citep{cohen_dragonfly_2018}, both confirming its spatial proximity to the host. ALFALFA likely adopts S09's group distance assignment $D_{\alpha}=11.1$ Mpc for most Leo I region low-mass objects, so the line-of-sight distance information is particularly less deterministic for evaluating association in this case (\S \ref{subsec:sat_search}). The remaining $2R_{200}$ object, UGC 5761 (NGC 3299), is a more massive dwarf galaxy loosely connected to the host. \citet{muller_leo-i_2018} summarizes the optical properties for many of the Leo Group dwarf galaxies (\S \ref{subsec:LeoI}; \S \ref{subsec:M66Group} below), including all the satellite candidates of M95.

NGC 3368 (M96) is a double-barred spiral galaxy with complex morphology, hosting a supermassive black hole in the central bulge \citep{erwin_double-barred_2004,2010MNRAS.403..646N}. As the primary of the Leo I Group, it has a similar mass to the MW ($M_{\rm Halo} \approx 1.3 \times 10^{12}$ $M_{\odot}$ from abundance matching), but is not an ideal MW analog because of the complex Leo I group environment. We assign five satellite candidates to M96, four of which have intermediate (AGC 200512=LeG 06, UGC 5812, AGC 202035) to high (AGC 202024=LeG 13, the only $R_{200}$ object of the five) likelihood of association. UGC 5812 is an Sdm galaxy \citep{karachentsev_updated_2013} in M95's projected $R_{200}$ field and outside of M96's, but is assigned to M96 because of a lower $\Theta$. The remaining candidate, UGC 6014, although identified as an M96 Group object by S09, is likely a background object given its large positive velocity separation and projected distance.

\subsection{NGC 3413}
NGC 3413 is a relatively isolated S0 galaxy and the least massive host in our sample. With a derived halo mass $M_{\rm Halo} \approx 6.3 \times 10^{10}$ $M_{\odot}$ and rotation velocity $V_{\rm rot} \approx 90$ km $\rm s^{-1}$ (Table \ref{table:analog_properties}), it is clearly not a MW analog, but the isolated environment indicates it is not a group member or satellite to larger systems. We consistently find no gaseous satellites in its vicinity across the different satellite selection methods.

\subsection{NGC 3486} \label{subsec:NGC3486}
NGC 3486 is an SAB(r)c galaxy hosting an x-ray faint active nucleus \citep{annuar_nustar_2020}, also classified as a Type 2 Seyfert \citep{ho_search_1997}. With a halo mass $M_{\rm Halo} \approx 4 \times 10^{11}$ $M_{\odot}$, it is of intermediate mass in our host sample, a factor of three less massive compared with the MW.

NGC 3486 appears unique in its abundance of gas-containing satellites (Figure \ref{fig:count_comprehensive}). It has six candidates within $R_{200}$ that are close to the host in both the velocity and position space with relatively high $F_{\rm d_{\rm 3D}}$. But we identify this high $R_{200}$ satellite number as a potential outlier because 4/6 of the $R_{200}$ candidates are missing clear optical counterparts -- AGC 208569, 722731, 219369, and 212945. 
We keep these sources in the satellite sample as potential dwarf galaxies unless future studies prove they are HI-only features. AGC 215232 (dI; \citealt{2015ApJS..217...27A}) and UGC 6102 ($M_{\rm HI} \approx 2.0 \times 10^{8}$ $M_{\odot}$ comparable to the SMC, LMC, and IC 10) are the two $R_{200}$ candidates verified as dwarf galaxies. UGC 6102's relatively low ${F_{\rm d_{\rm 3D}}}$ among the NGC 3486 satellite candidates (despite its kinematic proximity to the host) arises from the high local distance error $\sigma_{D}=4.2$ Mpc. Extending to $2R_{200}$, we find one more satellite: AGC 210027 (UGCA 225), a blue compact dwarf (BCD) with observed central star-forming regions \citep{cairos_multiband_2001,wakker_far_2003}.

\subsection{NGC 3593 and 3627 (M66 Group)}  \label{subsec:M66Group}
NGC 3593 and NGC 3627 (M66) have overlapping $R_{200}$ halos (Figure \ref{fig:virgo_field_sky_distribution}) and are at a derived three-dimensional distance of $\approx 547$ kpc. NGC 3593 is frequently associated with the Leo Triplet (M66) Group where M66 is the primary. The other two spiral members of the Triplet, NGC 3623 (M65) and NGC 3628, are located at $>10$ Mpc, and more than 1 Mpc away from the primary and are not included in our host sample due to the HI sensitivity limitations (\S \ref{sec:data}). Less than $1.5$ Mpc from the Leo Ring, the Leo Triplet field also contains HI-only structures (see, e.g., S09) that are excluded from our dwarf galaxy satellite sample. 

NGC 3593 is a starburst galaxy with a relatively centered star-forming HII region \citep{hunter_star_1989} classified as SA(s)0/a. Both satellite candidates shared between the virial halos of NGC 3593 and 3627, AGC 202256 and 210220 (IC 2684), are assigned to 3593 based on the lower $\Theta$ values. They both have relatively high $F_{\rm d_{\rm 3D}}$ among the satellite candidates, well-defined optical counterparts \citep{durbala_alfalfa-sdss_2020-1,muller_leo-i_2018}, and relatively low stellar masses $M_{*} < 10^{7}$ $M_{\odot}$.

NGC 3627 (M66) is the most massive spiral in our host sample. At $M_{\rm Halo} \approx 2. \times 10^{12}$ $M_{\odot}$, it is comparable to our adopted value for M31. All three of its $R_{200}$ objects, and the two we assign to NGC 3593, are confirmed by ELVES. Beyond $R_{200}$, there are three more $V_{\rm esc}$ candidates with low to intermediate $F_{\rm d_{\rm 3D}}$, and whose association is relatively unclear due to the complex group environment and potential contamination. 

\subsection{NGC 4274 and 4310 (NGC 4274 Group)} \label{subsec:NGC4274_ComaI}
NGC 4274 and 4310 is a pair with a complex local environment. They are often classified as members of the NGC 4274 (sub)group of $\sim 17$ members \citep{fouque_groups_1992,garcia_general_1993} that belong to the Coma I Group, which is part of a larger infalling group to Virgo \citep{1977ApJ...213..345G,2011MNRAS.412.2498M,kourkchi_galaxy_2017}. The two galaxies are placed at nearly the same 3D distance in the ALFALFA catalog, and their derived three-dimensional separation is only $154$ kpc. 

NGC 4274 is a double-barred spiral where the outer arms form a ring structure \citep{1995MNRAS.274..369S,erwin_double-barred_2004,2012ApJ...754...67F}. Significant non-convergence in literature distances challenges its mass estimation. As the literature distances range from $\sim 9.7$ Mpc to $\sim 19.4$ Mpc \citep{1988ngc..book.....T,tully_extragalactic_2009}, the ALFALFA value $D_{\alpha} = 9.6$ Mpc is on the low side; if, for example, $D_{\alpha}$ is a factor of $\sim 2$ underestimated, the current $M_{*} \approx 7.97 \times 10^{9}$ $M_{\odot}$ is a factor of $\sim 4$ underestimated, bringing the corrected value to a MW-like mass. This might explain why NGC 4274 has a relatively large HI FWHM $W_{50}=459$ km $\rm s^{-1}$ and a high derived $V_{\rm rot} \approx 248$ km $\rm s^{-1}$ (Table \ref{table:analog_properties} and Figure \ref{fig:analog_mass_stats}) despite an intermediate $M_{*}$. In addition, significant distance underestimation might result in the effective HI mass sensitivity decreasing from $\sim 10^{7}$ to $\sim 4 \times 10^{7}$ $M_{\odot}$ (Figure \ref{fig:survey_sensitivity_with_count}).

The satellite sample of NGC 4274 has large host-satellite velocity separations (all exceeding $V_{\rm esc}$, some approaching the 300 km $\rm s^{-1}$ upper limit), moderate to small spatial (projected and line-of-sight) separations, and large distance errors. We recognize that for all five $2R_{200}$ candidates, the large velocity differences lead to a very small likelihood of host-satellite gravitational association. The candidates are possibly foreground (AGC 724774 and UGC 7438) and background (UGC 7300, AGC 747848, 220408) contaminations. Future data in the NGC 4274 field will help constrain the local sensitivity/completeness and its satellite population.

NGC 4310 is a relatively faint spiral galaxy with an HI FWHM $W_{50}=155$ km $\rm s^{-1}$, close to the low-mass end of our host selection limit. We obtained $V_{\rm rot}=92$ km $\rm s^{-1}$ from an updated inclination angle measurement \citep{truong_high-resolution_2017}. Being distinctively less massive than the group primary, satellite candidates shared between NGC 4310 and 4274 are consistently assigned to the 4274 based on our group assignment method (equation \ref{eqn:dimensionless-separation}).

\subsection{NGC 4517} \label{subsec:NGC4517}
NGC 4517 (also denoted as NGC 4437) is an edge-on Sd galaxy. It declination $\delta \sim 0{\degr}$ places it at the edge of the ALFALFA Virgo field and hence it has partial sky coverage, though $>50 \%$ of the projected $R_{200}$ area is covered (Figure \ref{fig:virgo_field_sky_distribution}). It is an overlapping low-mass host between ALFALFA and ELVES, where our finding of no gaseous satellites within $R_{200}$ is confirmed by ELVES finding of no gaseous and one quenched (\S \ref{subsec:quenched_fraction}). Within $2R_{200}$, we find two candidates, AGC 225760 and 221570, with very low $F_{\rm d_{\rm 3D}}$ due to their $15-17$ Mpc ALFALFA assigned distances (Table \ref{table:2R200_satellites}; note 8a). We identify that the high ALFALFA satellite distances might result from velocity flow overestimates, and do not rule out potential satellite association, because both the host \citep{karachentsev_infall_2014} and the satellite AGC 221570 \citep{fouque_groups_1992,garcia_general_1993} are considered infalling galaxies into the Virgo cluster. In this single special case, the host's small $2R_{200}$ area (254 kpc, Table \ref{table:analog_properties}) is fully covered by ELVES, which within $2R_{200}$ identifies two gaseous satellites, including our AGC 221570, and three quenched. Their other gaseous satellite has a low stellar mass of $10^{5.98}$ $M_{\odot}$ and is likely below the ALFALFA detectability. Our other $2R_{200}$ object, AGC 225760, is rejected by ELVES due to a high SBF distance (13.19 Mpc, see their dw1229p0006). 

\subsection{NGC 4559}\label{subsec:NGC4559}
NGC 4559 is an intermediate SAB(rs)cd spiral in Coma Berenices with a range of distance measurements, and we adopt the TRGB distance of 8.91 Mpc \citep{2017AJ....154...51M} in replacement of the ALFALFA value $D_{\alpha} = 7.3$ Mpc. We find zero gaseous satellites in its $R_{200}$, but recognize that we miss a Leo T-like $M_{\rm HI} \approx 4 \times 10^{5}$ $M_{\odot}$ satellite located $\sim 0.4{\degr}$ ($\sim 58$ kpc) from its center \citep{vargas_halogas_2017} due to sensitivity. 

There is a drastic increase to 8 satellites when extending to $2R_{200}$ (Figure \ref{fig:count_comprehensive}), which we identify may result from contamination. Only 2/8 candidates, AGC 724993 (KUG 1236+293, dI; see \citealt{2015ApJS..217...27A}) and AGC 220847 (KUG 1234+299; see \citealt{karachentsev_morphological_2018}), are likely true satellites based on the velocity proximity and relatively high 3D confidence. Out of the remaining six candidates, contamination is likely high: four with intermediate separations are relatively uncertain, while the two with large positive velocity and 3D distance separations are likely background contaminants (Table \ref{table:2R200_satellites}; see note 8b).

\subsection{NGC 4826}\label{subsec:ngc4826}
NGC 4826 (M64) is the closest spiral host in our sample, which gives it a large projected area (Figure \ref{fig:virgo_field_sky_distribution}), and a better local HI detectability $M_{\rm HI}\approx 10^{6.2}$ $M_{\odot}$ (Figure \ref{fig:survey_sensitivity_with_count}). It is in a relatively isolated environment with few massive nearby galaxies \citep{1976ApJS...32..409T}. As an overlapping host between ALFALFA and ELVES, our $V_{300}$ object, AGC 742601 (IC 3840), is the only ELVES-identified gaseous satellite within $R_{200}$. The $V_{\rm esc}$ object (AGC 742601) being rejected by ELVES can be explained by the $F_{\rm d_{\rm 3D}}$ test using the updated distances \citep{2020ApJ...905..104K}, as described in \S \ref{appendix:Fd3d}. Extending to $2R_{200}$, we find AGC 221120 (equivalently F575-3 or KDG 215), a satellite with the highest 3D proximity confidence in the sample, $F_{\rm d_{\rm 3D}}=87.3\%$ (Table \ref{table:2R200_satellites}), located at $D=4.68\pm0.19$ Mpc \citep{2020ApJ...905..104K,schombert_anomalous_2021}. But the other $2R_{200}$ object, UGC 8030, is more likely a background field object given the large velocity and spatial separations.


\bsp	
\label{lastpage}
\end{document}